\documentclass[twocolumn,prd,aps,tightenlines,floats,floatfix,preprintnumbers,nofootinbib,eqsecnum]{revtex4-1}

\def\sp{\kern +3pt}
\def\sm{\kern -3pt}
\def\spQ{\kern +6pt}

\def\bea{\begin{eqnarray}}
\def\eea{\end{eqnarray}}
\def\etal{{\it et al.\/}}

\def\sfrac#1#2{{\textstyle \frac{#1}{#2}}}

\def\be{\begin{equation}}
\def\ee{\end{equation}}
\def\ba{\begin{eqnarray}}
\def\ea{\end{eqnarray}}

\usepackage{graphicx}
\usepackage{amsmath}
\usepackage{amssymb}
\usepackage{bbold}

\usepackage{xcolor}

\begin{document}

\phantom{0}
\vspace{-0.2in}
\hspace{5.5in}

\preprint{{\bf  LFTC-20-9/61}}

\vspace{-1in}

\title
{\bf Electromagnetic form factors of the $\Omega^-$ baryon in the spacelike and 
timelike regions}
\author{G.~Ramalho}
\vspace{-0.1in}

\affiliation{Laborat\'orio de 
F\'{i}sica Te\'orica e Computacional -- LFTC,
Universidade Cruzeiro do Sul and Universidade Cidade de  S\~ao Paulo,  \\
01506-000,   S\~ao Paulo, SP, Brazil}

\vspace{0.2in}
\date{\today}

\phantom{0}

\begin{abstract}
We present complete calculations of the electromagnetic form factors 
of the $\Omega^-$ in the spacelike region and in the timelike region.
The four elastic form factors: electric charge ($G_{E0}$), 
magnetic dipole  ($G_{M1}$),  electric quadrupole ($G_{E2}$) 
and magnetic octupole ($G_{M3}$), 
are estimated within the covariant spectator quark model,
in terms of the square momentum transfer $q^2$.  
The free parameters of the $\Omega^-$ wave function,
including a $S$-wave state and two independent $D$-wave states 
radial wave functions and the admixture coefficients 
are fixed by the comparison with the lattice QCD data 
in the spacelike region ($Q^2=-q^2 \le 0$) and with the recent 
$e^+ e^- \to \Omega^- \bar \Omega^+$ data from CLEO in 
the timelike region ($q^2 > 0$).
The estimates in the timelike region 
for square momentum transfer $q^2 \ge 4 M_\Omega^2$ 
are based on large-$q^2$ asymptotic relations
($M_\Omega$ is the $\Omega^-$ mass).
We examine also the impact of the large-$Q^2$ correlations between 
different form factors and analyze the possible solutions.
The electric quadrupole and the magnetic octupole moments of the $\Omega^-$, 
and the $e^+ e^- \to \Omega^- \bar \Omega^+$ integrated cross sections 
for very large $q^2$ are estimated based on the model results.
\end{abstract}

\vspace*{0.9in}  
\maketitle

\section{Introduction}
\label{secIntro}

The study of the electromagnetic structure of 
the $\Omega^-$, composed by three valence strange quarks, 
is very challenging.
Although the $\Omega^-$ is the more stable known baryon with spin 3/2
[longer mean life than the $\Delta(1232)$]
its physical properties are almost unknown,
apart the charge and the magnetic 
moment~\cite{Capstick00,Beg64,Muller,Omega,Omega2,Hyperons,Alexandrou10,Diehl91,Wallace95,PDG20,Afanasev12}.
[The mean life of the $\Omega^-$ is $8 \times 10^{-11}$ s, 
and the mean life of the $\Delta(1232)$ is $6\times 10^{-24}$ s].
The first measurement of the $\Omega^-$
{\it effective} form factor $|G(q^2)|$  
in the timelike region at CLOE
($e^+ e^- \to \Omega^- \bar \Omega^+$ reactions)~\cite{Dobbs14a}
opens a new window to probe the internal structure 
of the $\Omega^-$ and the properties of 
the form factors at large $|q^2|$.
The effective  form factor $|G(q^2)|$ is determined by 
a combination of the four $\Omega^-$ electromagnetic form factors.

Additional information about the $\Omega^-$ 
electromagnetic structure  
can be obtained from lattice QCD simulations, 
which can today be performed at the physical baryon mass 
in the spacelike region 
($Q^2 \ge 0$)~\cite{Boinepalli09,Alexandrou10}.
These simulations can be regarded as a good representation 
of the physical baryon because they are performed 
at the physical strange quark mass, 
and also because the effects of the meson cloud 
excitation of the baryon core are expected to be small
(heavy meson excitations are suppressed according with 
chiral perturbation theory)~\cite{NSTAR,NSTAR17}.

The measurement of the $\Omega^-$ electromagnetic form factors 
for finite $Q^2=-q^2$ is a hard task 
due to the difficulty in creating strange baryon targets 
which may be scattered by electron beams~\cite{NSTAR,Aznauryan12,Afanasev12}.
The long life of the  $\Omega^-$
can, however, be used to obtain 
accurate determinations of the 
magnetic moment~\cite{Diehl91,Wallace95,PDG20}.

A question that can be raised is 
how well we can estimate today 
the $\Omega^-$ electromagnetic form factors 
$G_{E0}$, $G_{M1}$, $G_{E2}$ and 
$G_{M3}$~\cite{Nozawa90,Alexandrou09,Alexandrou10,Pascalutsa07a,DeltaSFF,DeltaFF,Deformation}
in the different kinematic regions.
In the present work, we combine the present knowledge 
on the $\Omega^-$, including lattice QCD simulations 
in the spacelike region, measurements 
of the  $\Omega^-$ effective form factor in
the timelike region, and the expected analytic behavior 
for very large $|Q^2|$,
to shed some light on the dependence of the  
electric quadrupole and the magnetic octupole form factors on $Q^2$.
The constraints associated with the electromagnetic
form factors at the large $|Q^2|$ 
prove to have an important role in the shape of the form factors.
The timelike data are very pertinent to the process    
because they contain information about the \mbox{large-$|Q^2|$} region
and provide a unique test to the shape of the  
form factors in an extreme regime.
Although lattice QCD simulations can be used 
to infer the dependence of the form factors on $Q^2$,
they are presently limited in precision above $Q^2= 2$ GeV$^2$~\cite{Alexandrou10}.

The magnetic moment of the $\Omega^-$, the electric charge ($G_{E0}$), and the
magnetic dipole ($G_{M1}$) form factors have been estimated using 
several frameworks~\cite{Beg64,Omega,Bernard82,Tomozawa82,Georgi83,Krivoruchenko87,Kim89,Kunz90,Chao90,Schwesinger92,Gobbi92,Schlumpf93,Hong94,Butler94,Ha98,Linde98,Zhu98,Wagner00,Aliev00,Iqubal00,Kerbikov00,Franklin02,An06}.
Although there are a few estimates 
of the $\Omega^-$ quadrupole form factors~\cite{Gershtein81,Richard82,Isgur82,Krivoruchenko91,Leinweber92,Buchmann02,Arndt03,Ledwig09,Aliev09,Geng09b,Li17,Kim19},
including extrapolations from lattice QCD 
simulations~\cite{Alexandrou10,Omega2,Boinepalli09},
and of the  $\Omega^-$ octupole moment~\cite{Buchmann08,Aliev09,Nicmorus10,Sanchis13},
the information about the functions $G_{E2}$ and $G_{M3}$ is scarce.
Accurate lattice QCD simulations are at the moment limited 
to $0 \le Q^2 \le 2$ GeV$^2$.
The only available lattice QCD simulation for $G_{M3}$, gives 
$G_{M3} (0.23\; \mbox{GeV}^2) = 1.25 \pm 7.50$~\cite{Boinepalli09}.
There is then all the interest in studying the function $G_{M3}$, 
including the region near $Q^2=0$ and the respective
falloff with $Q^2$.
Experiments in facilities like 
BABAR~\cite{BaBar}, 
BES III~\cite{BESIII}, CLEO~\cite{Dobbs14a,Dobbs17a},
and PANDA~\cite{Singh17a} 
based on $e^+ e^-$ collisions
can also be used to access the electromagnetic structure
of the  $\Omega^-$~\cite{Dobbs17a,Hyperons}.

The $e^+ e^- \to B \bar B$ experiments,
where $B$ is a generic baryon, opens a new window 
to probe the electromagnetic structure of hyperons~\cite{Cabibbo61a,Dobbs14a},
hardly accessed in the spacelike region~\cite{Afanasev12}.
Of particular interest is the opportunity 
to study the correlations between different valence quark 
compositions, including quark pairs and 
others~\cite{Kroll93a,Jakob93b,Jaffe03,Wilczek04,Selem06,Dobbs14a,Dobbs17a}.
The first theoretical estimates of the $e^+ e^- \to B \bar B$ 
cross sections and hyperon effective form factor, in  the timelike region, 
were based on vector meson dominance (VMD) models~\cite{Korner77,Dubnickova93}.
More recently, with the emergence of accurate data for a variety of hyperons, 
new models have been proposed~\cite{Haidenbauer92,Liu17,Hyperons,Perotti19,Haidenbauer20}, 
including improved VMD 
models~\cite{Dalkarov10,Haidenbauer16,Faldt17,Cao18,Yang19,Li20a}.
Most of these studies focus on the $\Lambda$ and $\Sigma$ systems.
Theoretical studies of the $\Omega^-$ electromagnetic properties 
in the timelike region are rare~\cite{Perotti19,Hyperons,Dobbs14a,Dobbs17a}.

Our calculations of the  $\Omega^-$ electromagnetic form factors 
follow the formalism of the 
covariant spectator quark model~\cite{NSTAR17,Nucleon,NDelta,Nucleon2}
for spin-3/2 baryons.
In the formalism, the  $\Omega^-$  wave function 
is represented by a combination of a dominate 
$S$-wave state and two $D$-wave states~\cite{NDeltaD,DeltaFF,Omega,Omega2}.
The mixture parameters and the radial structure 
of the three components are determined by fits to the available data 
(the magnetic moment, lattice QCD, and effective timelike form factor) 
as well by the 
expected behavior of the form factors for very large $Q^2$.
At large $Q^2$, we consider also a relation between 
the form factors $G_{M1}$ and $G_{M3}$, derived from the asymptotic 
behavior of the helicity transition amplitudes at large $Q^2$.
At the end, we use our best parametrization to 
make predictions to the electric quadrupole 
and magnetic octupole moments of the $\Omega^-$
and the effective form factor $|G(q^2)|$ at large $q^2$.

Although the present analysis is dominated by spacelike data,
we conclude that the information about the function $G_{M3}$
and the effective form factor $|G (q^2)|$ 
is important to determine the shape of the form factors.
The timelike and $G_{M3}$ data are represented by three points,
while the remaining data (lattice QCD) are represented by 100 points. 
We conclude also that accurate 
lattice QCD calculations of $G_{M3}$, 
possible with the present state-of-the-art methods,
and measurements of the $e^+e^- \to \Omega^- \bar \Omega^+$
cross sections at large $q^2$ can further help to infer the shape 
of the form factors at large  $|Q^2|$ and 
to reduce the uncertainty of $G_{M3}(0)$.

The present article is organized as follows.
In the next section, we discuss in detail
the available experimental and theoretical 
information about the $\Omega^-$ form factors.
In Sec.~\ref{secCSQM}, we discuss the covariant 
spectator quark model and the formalism associated 
to baryons with spin 3/2 and positive parity.
The calculations of the $\Omega^-$ electromagnetic 
form factors in the spacelike region ($Q^2 \ge 0$)
are presented in Sec.~\ref{secModel}.
The extension of the model for the timelike region 
($q^2= -Q^2 > 0$) and our final results are presented 
and discussed in  Sec.~\ref{secOmega-TL}.
In Sec.~\ref{secConclusions}, we present the outlook and conclusions.

\section{Electromagnetic structure of the $\Omega^-$ baryon}
\label{secOmega}

The $\Omega^-$ is a baryon with spin 3/2 and positive parity ($J^P= \frac{3}{2}^+$).
As a consequence the transition current is characterized 
by four independent structure functions
dependent on $Q^2$~\cite{Nozawa90,Pascalutsa07a,Alexandrou09,Alexandrou10}.
The most common representation of those 
structure functions is the multipole form factor representation, 
where the $\Omega^-$ structure is described 
by the electric charge ($G_{E0}$), magnetic dipole ($G_{M1}$),
electric quadrupole ($G_{E2}$) 
and magnetic octupole ($G_{M3}$) form factors~\cite{Omega2,DeltaSFF,DeltaFF,Alexandrou10}.
The definition of the multipole form factors 
is presented in Appendix~\ref{app-PQCD}.
The electric charge and the magnetic dipole 
form factors provide information about the distribution of 
charge and magnetism inside the baryons.
The electric quadrupole and magnetic octupole 
measure the deviations from the distributions 
from a symmetrical form ($G_{E2} \ne 0$ and $G_{M3} \ne 0$), providing 
a direct evidence of the deformation of 
the baryons~\cite{Deformation,Alexandrou10,Alexandrou09,Buchmann02,Buchmann08,Buchmann01}.

As pointed out already, except for the electron-positron 
collisions~\cite{Dobbs14a,Hyperons},
the $\Omega^-$ baryon is difficult to produce 
in the laboratory~\cite{Diehl91,Wallace95,Afanasev12},
due the structure based on three strange quarks.
For this reason the electromagnetic structure of 
the  $\Omega^-$ is almost unknown, except for the charge ($-e$)
and the magnetic moment $\mu_{\Omega}$.

We review next our sources of information about the
$\Omega^-$ electromagnetic structure.

\subsection{Experimental data}

The long lifetime ($\tau_\Omega \simeq 8 \times 10^{-11}$ s, 
decay by weak interaction)~\cite{PDG20}
allows a precise determination of the  $\Omega^-$ magnetic moment.
The  $\Omega^-$ magnetic moment has been measured 
a few times with different precisions~\cite{Diehl91,Wallace95}.
The Particle Data Group~(PDG) presents the world's average: 
$\mu_{\Omega} = (-2.02 \pm 0.05) \mu_N$, in nucleon magnetons 
($\mu_N \equiv \frac{e}{2 M_N}$, 
where $M_N$ is the nucleon mass
and $e$ is the elementary charge). 
In the present work, we use the result from PDG~\cite{PDG20} 
corresponding to 
\ba
G_{M1}(0) = -3.60 \pm 0.09, 
\label{eqGM10}
\ea
based on $\mu_\Omega \equiv G_{M1} (0) \frac{e}{2 M_\Omega}$.

\subsection{Results from lattice QCD simulations}

Since the structure of the $\Omega^-$ is dominated by 
three strange valence quarks, one assumes that 
the electromagnetic form factors can in a  
good approximation be simulated by lattice QCD calculations 
at the strange quark physical mass.
These simulations are possible in the present days,
as shown in several lattice QCD simulations~\cite{Alexandrou10,Aubin09}.
Although it may be argued that the valence quark structure 
is not the complete picture, and that sea quark effects 
must also be taken into account,
it is known that those effects are dominated by 
the kaon cloud, since the pions cannot be produced directly by 
three strange quark cores. 
The kaon and eta cloud effects, however, are suppressed 
according to chiral perturbation theory~\cite{Jenkins91,Meissner97,Bernard08}.
The conclusion is then that the lattice QCD simulations 
at the physical strange quark mass can be 
interpreted as an accurate simulation 
of the physical results, and 
that no extrapolation of the results is necessary
The only limitation of these calculations 
is the intrinsic errors associated with lattice QCD simulations,
such as the size of the lattice spacing and the finite volume 
of the simulations.

In this work, we consider the more consistent 
simulation of the  $\Omega^-$ electromagnetic form factors 
from Alexandrou \etal~\cite{Alexandrou10}, 
as a reliable representation of the  
$\Omega^-$ physical electromagnetic form factors.
Early lattice QCD simulations can be found 
in Refs.~\cite{Bernard82,Leinweber92}.

The simulations from Ref.~\cite{Alexandrou10} are based 
on two unquenched methods: the domain-wall fermions (DWF) method
and the hybrid action method.
The simulations from  Ref.~\cite{Alexandrou10} are 
restricted to the form factors $G_{E0}$, $G_{M1}$ and $G_{E2}$.
These simulations provide data for $Q^2$ up to 4 GeV$^2$,
but only the data for $Q^2 \le 2$ GeV$^2$ are relatively precise.
The hybrid action simulations correspond to $m_\pi= 0.353$ MeV.
For the DWF there are simulations for 
$m_\pi = 297$, 330 and 355 MeV.
Ideally, we should select the simulations corresponding 
to the lower pion mass for each method.
Unfortunately, some datasets do not include results 
for all form factors or the data statistic are poor.
To obtain the complete picture of the 
form factors $G_{E0}$, $G_{M1}$, and $G_{E2}$,
in this work, we use then the four sets 
of lattice QCD data from Ref.~\cite{Alexandrou10}.

The only direct information about 
the octupole  magnetic form factors come from  
lattice QCD simulations from Boinepalli \etal~\cite{Boinepalli09}
for the $\Delta^-$ form factors in the $SU(3)$ limit, 
when the pion mass is $m_\pi = 0.697$ GeV.
In this point the $d$ quarks and $s$ quarks have the same properties 
and masses, and the $\Delta^-$ structure resembles the $\Omega^-$ structure.
The simulation from  Ref.~\cite{Boinepalli09} 
overestimates the physical $\Omega^-$ mass.
It provides, nevertheless, the only available estimate 
of the form factor $G_{M3}$ based on QCD first principles.
These estimates are performed at one single point $Q^2=0.23$ GeV$^2$.
The result of the magnetic octupole form factor is 
$G_{M3} (Q^2)= 1.25 \pm 7.50$~\cite{Boinepalli09}.

\subsection{$\Omega^-$ data in the timelike region}

In recent years there have been 
important experimental developments 
in the study of the baryon structure in the timelike region, 
based on the electron-positron collisions 
in facilities like BABAR, BES III and 
CLEO~\cite{Pacetti15a,BaBar,Dobbs14a,Dobbs17a,BESIII}.
From the $e^+ e^- \to B \bar B$ reactions one has access 
to the electromagnetic structure of the baryon $B$,
in the region $Q^2 =- q^2 < 0$.
The threshold of the transition is $q^2 =-Q^2 \ge 4 M_B^2$
($M_B$ is the baryon mass).
Of particular interest has been the production 
of $\Omega^-$ baryon and the respective antistate 
in CLEO~\cite{Dobbs17a}.

In those experiments the integrated  $e^+ e^- \to B \bar B$ 
cross section in the $e^+ e^- $ center-of-mass frame  
becomes~\cite{Pacetti15a,Hyperons,Dobbs17a}
\ba
\sigma (q^2) = \frac{4 \pi \alpha^2 \beta C}{3 q^2} 
\left( 1 + \frac{1}{2 \tau_{\scriptscriptstyle T}} \right) |G(q^2)|^2, 
\ea 
where $G(q^2)$ is an effective form factor, 
$\tau_{\scriptscriptstyle T} = \frac{q^2}{4 M_B^2}$, 
\mbox{$\alpha \simeq \frac{1}{137}$}
is the fine-structure constant, $\beta$ is a kinematic factor 
defined by $\beta = \sqrt{1 - \frac{1}{\tau_{\scriptscriptstyle T}}}$,
and $C$ is a factor which depends 
on the charge of $B$~\cite{Pacetti15a,Hyperons}.
For large $q^2$, one has $C \simeq 1$.

The effective form factor $G (q^2)$ for 
baryons with spin 1/2 and positive parity ($J^P=\frac{1}{2}^+$ states) 
takes the form~\cite{Dobbs14a,BaBar,Tzara70a,Denig13,Haidenbauer14}
\ba
|G (q^2)|^2 &=& 
\left( 1 + \frac{1}{2 \tau_{\scriptscriptstyle T}} \right)^{-1}
\left[  
|G_M (q^2)|^2 + \frac{1}{2 \tau_{\scriptscriptstyle T}} |G_E (q^2)|^2\right], \nonumber \\
& =& 
\frac{2 \tau_{\scriptscriptstyle T} |G_M (q^2)|^2  
+ |G_E (q^2)|^2 }{2 {\tau_{\scriptscriptstyle T}} + 1},
\label{eqGeff1}
\ea
where $G_E$ and $G_M$ are the electric charge and magnetic dipole form factors.

Equation (\ref{eqGeff1}) is still valid for $\frac{3}{2}^+$ states, 
if we use the replacements~\cite{Korner77}
\ba
& &
|G_E|^2 \to 2 |G_{E0}|^2 + \frac{8}{9} (\tau_{\scriptscriptstyle T})^2 |G_{E2}|^2, 
\label{eqGE2abs} \\
& &
|G_M|^2 \to \frac{10}{9}|G_{M1}|^2 + \frac{32}{5}  (\tau_{\scriptscriptstyle T})^2 |G_{M3}|^2.
\label{eqGM2abs}
\ea

\subsection{Perturbative QCD constraints}

Additional information about the $\Omega^-$ form factors 
come from perturbative QCD (pQCD)
for very large $Q^2$ (or $q^2)$~\cite{Carlson0,Carlson,Brodsky}.
As for the case of the nucleon, where pQCD estimates 
show that $G_{E}$, $G_M \propto 1/Q^4$,
also in the case of $\frac{3}{2}^+$ baryons, 
one can estimate the falloff of the form factors 
for very large $Q^2$.
For the $\frac{3}{2}^+$ resonances,  
one obtains
\ba
& &
G_{E0}, \; G_{M1} \propto \frac{1}{Q^4}, 
\label{eqGE0lQ2}\\
& &
G_{E2}, \; G_{M3} \propto \frac{1}{Q^6},
\label{eqGE2lQ2}
\ea
apart logarithmic corrections~\cite{Carlson0,Brodsky}.
[Meaning that the leading-order dependence can 
include factors $\log (Q^2)$, or powers of $\log (Q^2)$,
which are negligible in comparison with $Q= \sqrt{Q^2}$].
The corollary of these results is that $G(q^2) \propto 1/q^4$ 
for very large $q^2$,
as a consequence of the asymptotic relations between 
spacelike and timelike form factors~\cite{Hyperons}.
Those relations are discussed in Sec.~\ref{secSLTLfit}.

The analysis of the helicity transition
amplitudes at large $Q^2$ imposes, however,
a constraint stronger than (\ref{eqGE0lQ2}) and (\ref{eqGE2lQ2}).
From the study of the asymptotic behavior 
of the helicity transition amplitudes~\cite{Carlson0,Carlson}, one concludes 
that the magnetic-type 
form factors are related for very large $Q^2$ by
\ba
& &
G_{M1} = \frac{4}{5} \, \tau \, G_{M3},
\label{eqGM3extra}
\ea
where $\tau = \frac{Q^2}{4M^2}$.
In this notation $\tau = - \tau_{\scriptscriptstyle T}$.
The previous relation is derived in Appendix~\ref{app-PQCD}.
The error expected in the relation is terms of the order of $1/Q^6$.

The condition (\ref{eqGM3extra})
may look surprising at first.
One needs to keep in mind, however,  
that correlations between transition form factors at large $Q^2$
are common on electromagnetic transitions between baryon states.
Examples are some $\gamma^\ast N \to N^\ast$ transitions,
when $N^\ast$ are $\frac{3}{2}^+$ and $\frac{3}{2}^-$ states,
as the $\Delta(1232)$ and the $N(1520)$.
In those cases one has $G_M \simeq - G_E$, 
for very large $Q^2$~\cite{Devenish76,Compton,NDeltaD,N1520}.
Although those relations are related to the falloff of the 
transverse transition amplitudes 
($A_{1/2}$ and $A_{3/2}$)~\cite{Siegert2},
those constraints are only taken into account implicitly 
in some quark models and 
in some parametrizations of the data~\cite{Compton}.
The condition (\ref{eqGM3extra}) is also valid for 
the $\Delta(1232)$ elastic form factors~\cite{DeltaSFF,DeltaFF}
and for the other decuplet baryon members.

As far as we know, the constraint (\ref{eqGM3extra})
has not been discussed in the literature,
but it has a significant impact on our final results 
for the $\Omega^-$ electromagnetic form factors.
The relation (\ref{eqGM3extra}) is, however, 
the consequence of the natural order 
of the transition amplitudes between $\frac{3}{2}^+$ and 
$\frac{3}{2}^+$ baryon states (see Appendix~\ref{app-PQCD}).


\section{Covariant Spectator Quark Model}
\label{secCSQM}

In the present section, we discuss the formalism
associated with the 
covariant spectator quark model~\cite{Nucleon,Omega,NSTAR17}.
The model was developed within the covariant 
spectator theory~\cite{Gross}.
In the framework the baryons are interpreted 
as systems of three-constituent quarks 
where a quark is free to interact with electromagnetic probes
in relativistic impulse approximation~\cite{Nucleon,NDelta,Omega}.
Integrating over the degrees of freedom of the
non-interacting quarks, one reduces the three-quark system 
to a quark-diquark system where the spectator quark pair 
is represented by an on-mass-shell diquark with 
an average mass $m_D$~\cite{Nucleon,Omega,Nucleon2}.
One obtains then an effective quark-diquark wave function, 
free of singularities which describe the quark confinement 
implicitly~\cite{Nucleon,Nucleon2}.

The wave functions of the baryons are built according 
to the spin-flavor-radial symmetries where 
the radial wave functions are determined phenomenologically 
by the experimental data or by lattice QCD data 
for some ground state 
systems~\cite{NSTAR17,NSTAR,NDeltaD,LatticeD,Omega,OctetFF1}.
In the electromagnetic interaction with the quarks, 
we take into account the structure related 
to the gluon and quark-antiquark dressing.
To parametrize this structure we use a form based on VMD to 
represent the constituent quark electromagnetic
form factors~\cite{Nucleon,Omega,OctetFF1}.

The formalism has been applied extensively
to the study of the electromagnetic structure of 
several baryons in the 
spacelike region ($Q^2 \le 0$)~\cite{Nucleon,Nucleon2,Omega2,NDelta,NDeltaD,OctetFF1,OctetDecuplet,N1520,Lattice,Roper,SRapp,N1535,Siegert3}
and in the timelike region 
($Q^2 < 0$)~\cite{Hyperons,OctetDecupletTL,NDeltaTL,Timelike}.
The formalism has also been used in the study of 
the spacelike electromagnetic form factors of baryons  
in the lattice QCD regime~\cite{Lattice,LatticeD,Omega,OctetFF1} 
and in the nuclear medium~\cite{OctetFF2}.

\subsection{Transition current}

In the relativistic impulse approximation, 
the transition current between two baryon states, $B$ and $B'$,
described by quark-diquark wave functions, 
$\Psi_{B'}$ and $\Psi_B$, takes the form~\cite{Nucleon,Omega,Nucleon2}
\ba
J_{B',B}^\mu 
= 3 \sum_{\Gamma} 
\int_k \overline \Psi_{B'}(P_+,k) j_q^\mu  \Psi_{B}(P_-,k),
\label{eqJ-total}
\ea
where $P_+$, $P_-$, and $k$ are the final, initial and diquark momenta; 
$j_q^\mu$ is the quark current operator; and $\Gamma$ 
labels the diquark scalar and vector components.
The factor 3 takes into account the contributions 
associated with the different diquark pairs.
The integral symbol represents the covariant 
integration on the on-shell diquark momentum.

When we include the explicit form of 
the wave functions $\Psi_{B'}$ and $\Psi_B$,
we reduce $J^\mu_{B',B}$ to a Lorentz-invariant form 
projected into the asymptotic states of $B'$ and $B$.
Each gauge-invariant term defines 
an independent form factor.
For details about the $\frac{3}{2}^+$ 
elastic form factors check 
Refs.~\cite{DeltaSFF,DeltaFF,Nozawa90,Pascalutsa07a,Alexandrou09}
and Appendix~\ref{app-PQCD}.

In the following, we consider the elastic case ($B'= B$), 
since our focus is the electromagnetic form factors 
of $\frac{3}{2}^+$ baryons.

The quark current operator has the generic form
\ba
j_q^\mu = j_1 (Q^2) \gamma^\mu + j_2(Q^2) \frac{i \sigma^{\mu \nu} q_\nu}{2 M_N}
\label{eqJqMU}
\ea
where $M_N$ is the nucleon mass, as before,
and $j_1$ and $j_2$ 
are the Dirac and Pauli $SU (3)$ flavor operators,
respectively.
Equation (\ref{eqJqMU}) was defined for 
the first time for the study 
of the nucleon elastic form factors~\cite{Nucleon}.
The quark current $j_q^\mu$ was later 
extended to baryons 
with strange quarks~\cite{Omega,OctetFF1,OctetFF2,Hyperons}.

The operators $j_i$ ($i=1,2$) can be 
decomposed as 
\ba
j_i (Q^2) = 
\frac{1}{6} f_{i+} (Q^2)\lambda_0  +
\frac{1}{2} f_{i-} (Q^2)\lambda_3 +
\frac{1}{2} f_{i0} (Q^2)\lambda_s,
\nonumber \\
\label{eqJi}
\ea
where
\ba
&  
\lambda_0=\left(\begin{array}{ccc} 1&0 &0\cr 0 & 1 & 0 \cr
0 & 0 & 0 \cr
\end{array}\right), \hspace{.3cm}
& \lambda_3=\left(\begin{array}{ccc} 1& 0 &0\cr 0 & -1 & 0 \cr
0 & 0 & 0 \cr
\end{array}\right),
\nonumber \\
& 
\lambda_s = \left(\begin{array}{ccc} 0&0 &0\cr 0 & 0 & 0 \cr
0 & 0 & -2 \cr
\end{array}
\right)   &,
\label{eqL1L3}
\ea
are the flavor operators acting on the quark wave function in 
the flavor space,
$q=  (\begin{array}{c c c} \! u \, d \, s \!\cr
\end{array} )^T$. 
The functions  $f_{i+}$, $f_{i-}$ ($i=1,2$) 
represent the quark isoscalar and isovector 
form factors, respectively, based on 
the combinations of the quarks $u$ and $d$~\cite{Nucleon}.
The functions $f_{i0}$ ($i=1,2$) represent 
the structure associated with the strange quark~\cite{Omega}.  

The quark isoscalar and isovector (light) form factors,
are important for the study of the nucleon, 
the octet baryon, the decuplet baryon and the 
transitions between the octet baryon and 
decuplet baryon~\cite{Nucleon,Omega,OctetFF1,OctetFF2,OctetDecuplet}
but are not relevant to the present work.

When we consider a baryon composed 
exclusively of strange quarks, like the $\Omega^-$ baryon,
only the terms in $f_{i0}$ survive 
when we project $j_i$ into the flavor wave functions.

To parametrize the strange the strange quark form factors, 
we use the form inspired by the VMD mechanism~\cite{Omega}
\ba
& &
\hspace{-1.5cm}
f_{1 0} = \lambda_q
+ (1-\lambda_q)
\frac{m_\phi^2}{m_\phi^2+Q^2} + c_0 \frac{M_h^2 Q^2}{(M_h^2+Q^2)^2}, 
\label{eqQff1} \\
& &
\hspace{-1.5cm}
f_{2 0} = \kappa_s
\left\{
d_0  \frac{m_\phi^2}{m_\phi^2+Q^2} + (1-d_0)
\frac{M_h^2}{M_h^2+Q^2}  \right\},
\label{eqQff2}
\ea
where $m_\phi$ and $M_h$ are the vector meson masses,  
corresponding, respectively,
to the light vector meson 
($\phi$ meson, associated with an $s \bar s$ state)
and an effective heavy meson with mass $M_h= 2 M_N$, 
which simulate the short-range phenomenology.
The parameter $\lambda_q$ is determined 
by the study of deep inelastic scattering~\cite{Nucleon},
$\kappa_s$, $c_0$, and  $d_0$ are determined by the study of 
the decuplet baryon electromagnetic form factors~\cite{Omega},
based on the lattice QCD simulations from Ref.~\cite{Boinepalli09}.
The calibration of the strange quark form factors
takes into account also the experimental value 
for the $\Omega^-$ magnetic moment~\cite{Omega}.
The numerical values of the free parameters 
are $\lambda_q= 1.21$, $\kappa_s=1.462$, 
$c_0= 4.427$ and $d_0=-1.860$.

\subsection{Wave functions of spin-$\frac{3}{2}^+$ baryons}

We now review the formalism associated 
with the $\frac{3}{2}^+$ baryon states,
developed in previous works in the study 
of the $\Delta(1232)$ and the $\Omega^-$ 
systems~\cite{DeltaSFF,DeltaFF,NDelta,NDeltaD,Omega,Omega2,LatticeD}. 
We assume that the state corresponds 
to the $\frac{3}{2}^+$ baryon ground state 
(no radial excitations). 
The differences to the previous works are in 
the flavor states ($\Omega^-$ system) and 
in the radial wave functions.

We can decompose the wave functions of the $\frac{3}{2}^+$ baryon $B$ 
into three main components, 
associated with a mixture of an $S$ state and 
two $D$-states, labeled here as $D3$ and $D1$ states,
for the quark-diquark relative motion~\cite{Omega2}
\ba
\Psi_B (P,k)=
N \left[ 
\Psi_S(P,k)+ a \Psi_{D3}(P,k) + b \Psi_{D1}(P,k) 
\right], \nonumber \\
\label{eqPsiTotal}
\ea
where $a$ and $b$ are the $D$ state mixture coefficients 
of the states $D3$ and $D1$, respectively, 
and $N$ is the normalization constant
(assuming that the individual states are properly normalized).
The state $D3$ describes the configuration 
where the sum of the spin of the three quarks is $3/2$.
The state $D1$  describes the configuration 
where the sum of the spin of the three quarks is $1/2$.

The interpretation of the states as $S$- 
and $D$-wave components comes from the structure 
of the states in the rest frame.
In a moving frame, the intrinsic $S$- and $D$-wave 
states are modified, and other partial 
waves are generated~\cite{NDelta,NDeltaD}.

In the following, we refer states of core-spin ${\cal S}$  
to refer to states where the sum of the spin of the three quarks is ${\cal S}$
(ignoring the relative angular momentum).
The possible states for systems of three quarks
are then  ${\cal S}=1/2$ or ${\cal S}=3/2$.

In the present study we are not taking into 
account contributions associated with $P$-wave states.
Those contributions may be relevant for the nucleon 
and the octet baryon~\cite{Nucleon2,Axial}
but appear to not be so relevant for the $\frac{3}{2}^+$ states.
Notice that the $\gamma^\ast N \to \Delta(1232)$ 
transition can be described accurately by a combination 
of $S$ and $D$ states on the $\Delta(1232)$ 
wave function~\cite{NDeltaD,LatticeD,Siegert3}.

The explicit expressions for the $S$, $D3$, and $D1$ 
components of the wave functions 
are presented next.

\subsubsection{$S$-state wave function}
   
The $S$-state contribution to the wave function of 
the baryon $B$,
corresponding to a quark-diquark system 
with relative angular momentum $L=0$, 
can be written as~\cite{NDelta,NDeltaD,Omega}
\ba
\Psi_S(P,k) = 
- \psi_S(P,k) \left| B \right>_F 
\varepsilon_{\lambda P}^{\alpha *} u_\alpha (P,s),
\label{eqPsiS}
\ea
where $\psi_S(P,k)$ is the $S$ state radial wave function,
$\left| B \right>_F$ is the baryon $B$ flavor wave function,
$\varepsilon_{\lambda P}^{\alpha *}$ is the 
diquark polarization state ($\lambda=0,\pm$) 
in the fixed-axis basis~\cite{Nucleon,FixedAxis},
$u_\alpha (P,s)$ is the 
Rarita-Schwinger spinor~\cite{Rarita41,Benmerrouche89},
and $s$ is the spin projection of the baryon.
The indices $\lambda$ and $s$ are omitted 
on $\Psi_S(P,k)$ for simplicity.

Equation (\ref{eqPsiS})  generalizes the non relativistic structure 
of a three-quark wave function of 
the $\frac{3}{2}^+$ ground state baryon and satisfies the Dirac equation
\mbox{$({\not \! P} -M_B) \Psi_S(P,k)=0$}~\cite{NDelta,FixedAxis,NDeltaD}.

\subsubsection{$D$-state wave functions}

The construction of the states associated to 
quark-diquark configurations with a relative angular momentum $L=2$
requires the derivation of a ${\cal D}$-state operator 
and also a consideration of projectors ${\cal P}_{1/2}$ 
and ${\cal P}_{3/2}$, which decompose generic states 
into their components into states of core-spin 1/2 
and states of core-spin 3/2, respectively.

The ${\cal D}$-state operator can be expressed in terms of 
the momentum~\cite{NDeltaD} 
\ba
\tilde k^\alpha = k^\alpha - \frac{P \cdot k}{M_B^2} P^\alpha,
\ea
which can be used for the initial diquark 
($P=P_-$ and $\tilde k \to \tilde k_-$) 
or the final diquark ($P=P_+$ and $\tilde k \to \tilde k_+$).
At the baryon rest frame $\tilde k= (0,{\bf k})$.
Using this notation we can define the 
${\cal D}$-state operator~\cite{NDeltaD} as
\ba
{\cal D}^{\alpha \beta} (P,k) =
\tilde k^\alpha \tilde k^\beta - \frac{1}{3} \tilde k^2 \tilde g^{\alpha \beta},
\label{eqD-operator}
\ea
where
\ba
\tilde g^{\alpha \beta} = g^{\alpha \beta} - \frac{P^\alpha P^\beta}{M_B^2}.
\label{eq-Gtilde}
\ea
Note that Eqs.~(\ref{eqD-operator}) and (\ref{eq-Gtilde})
can be defined in the initial state ($P_-, \tilde k_-$) 
or in the final state  ($P_+, \tilde k_+$).

To separate the states of core-spin 1/2 
from  the states of core-spin 3/2, 
we consider the two projectors:
\ba
& &
({\cal P}_{3/2})^{\alpha \beta} = \tilde g^{\alpha \beta}
- \frac{1}{3} \tilde \gamma^\alpha \tilde \gamma^\beta, 
\label{eqP32} \\
& &
({\cal P}_{1/2})^{\alpha \beta} = \frac{1}{3} 
\tilde \gamma^\alpha \tilde \gamma^\beta, 
\label{eqP12} 
\ea 
where 
\ba
\tilde \gamma^\alpha = \gamma^\alpha - \frac{{\not \! P} P^\alpha}{M_B^2}.
\ea
The properties of these projectors 
are known in the literature~\cite{Benmerrouche89,NDelta,NDeltaD}.

To represent the two $D$ states in a compact form 
it is convenient to define also the state
\ba
W^\alpha (P,k;s) = {\cal D}^{\alpha \beta} (P,k) \, u_\beta(P,s).
\ea
One obtains two different states, $D3$ and $D1$,  
when we use the core-spin projectors 
${\cal P}_{3/2}$ and ${\cal P}_{1/2}$, defined by 
Eqs.~(\ref{eqP32}) and (\ref{eqP12})
\ba
W_{D3}^\alpha (P,k;s) = 
({\cal P}_{3/2})^{\alpha \beta} \, W_\beta(P,k;s), \\
W_{D1}^\alpha (P,k;s) = 
({\cal P}_{1/2})^{\alpha \beta} \, W_\beta(P,k;s).
\ea

The wave functions of the states $D3$ and $D1$ 
can now be written as~\cite{NDeltaD,DeltaFF}
\ba
& &
\Psi_{D3} (P,k) = 
- 3\; \psi_{D3} (P,k) \left| B \right>_F 
(\varepsilon_{\lambda P}^{*})_\alpha \, W_{D3}^\alpha (P,k;s), \nonumber \\
&& \label{eqPsiD3}\\
& &
\Psi_{D1} (P,k) = 
- 3\; \psi_{D1} (P,k) \left| B \right>_F 
(\varepsilon_{\lambda P}^{*})_\alpha \,  W_{D1}^\alpha (P,k;s), \nonumber \\
&& \label{eqPsiD1}
\ea
where $\psi_{D3}$ and  $\psi_{D1}$ are 
the $D3$ and $D1$ radial wave functions, respectively,
and the factor $-3$ was included by convenience
in order to mimic the form of $\Psi_S$ from (\ref{eqPsiS}) 
and to simplify the normalization condition 
of the two $D$ states~\cite{NDeltaD}.
As for the $S$ state, the wave functions 
$\Psi_{D3} (P,k)$ and $\Psi_{D1} (P,k)$ 
are both solutions of the Dirac equation. 

The normalization of the radial wave functions 
is discussed in the next section 
[see Eqs.~(\ref{eqNorma})], along with 
the discussion of the form of the radial wave functions.

One can demonstrate that the states $u^\alpha(P,s)$,
$W_{D3}^\alpha(P,k,s)$ are states with core-spin 3/2,
since the projection with ${\cal P}_{1/2}$ is zero, 
and the states are unchanged by the projector 
${\cal P}_{3/2}$~\cite{NDeltaD}.
As for the state $W_{D1}^\alpha(P,k,s)$, it is
a state with core-spin 1/2, since the projection with 
${\cal P}_{3/2}$  is zero, and it remains 
unchanged when projected by ${\cal P}_{1/2}$~\cite{NDeltaD}.
Furthermore, it was proved that Eqs.~(\ref{eqPsiD3})
and (\ref{eqPsiD1}) generalize the non relativistic 
wave function of three-quark $D$ states 
ground state  with core-spin 3/2 and 1/2, respectively~\cite{NDeltaD}.

\section{Spacelike model for the $\Omega^-$ baryon}
\label{secModel}

We discuss now the results of the covariant spectator quark model
for the  $\Omega^-$ electromagnetic form factors 
in the spacelike region.
We consider the  $\Omega^-$ wave function 
described by a combination of an $S$,
a $D3$, and a $D1$ state for a  
$\frac{3}{2}^+$ ground-state baryon, 
as in Eq.~(\ref{eqPsiTotal}) 
with the flavor state
$ \left| B \right>_F =  \left| sss \right> $.
In the following, we replace $B$ by $\Omega$ in the 
wave functions and masses.

The explicit expressions for the  $\Omega^-$ electromagnetic form factors 
are derived in Refs.~\cite{Omega2,NDeltaD}.
The final results depend on the 
parametrization of the strange quark form factors (\ref{eqQff1}) 
and (\ref{eqQff2}) 
and on the form of the radial 
wave functions $\psi_S$, $\psi_{D3}$ and $\psi_{D1}$.

Before presenting the final expressions to
the electromagnetic form factors, we
discuss the parametrizations to the 
radial wave functions.

\subsection{Radial wave functions}

Following the formalism of the covariant spectator quark model, 
we express the radial wave functions in terms 
of the dimensionless variable
\ba
\chi = \frac{(M_\Omega -m_D)^2 - (P-k)^2  }{M_\Omega m_D}.
\ea
This representation is justified in the cases
that the baryons and the diquark are both on-mass-shell~\cite{Nucleon,NSTAR17}.

For the  $S$, $D3$, and $D1$ states,  
we consider the radial wave functions~\cite{Omega},
\ba
& & \psi_S (P,k) = \frac{N_S}{m_D(\alpha_1 + \chi) (\alpha_2 + \chi)}, 
\label{eq-psiS}\\
& & \psi_{D3} (P,k) = \frac{N_{D3}}{m_D^3(\alpha_3 + \chi)^4}, 
\label{eq-psiD3}
\\
& & \psi_{D1} (P,k) = \frac{N_{D1}}{m_D^3(\alpha_4 + \chi)^4},
\label{eq-psiD1}
\ea
where $N_S$, $N_{D3}$, and $N_{D1}$ are normalization constants
and $\alpha_i$ ($i=1,..,4$) are square momentum range 
parameters in units $M_\Omega m_D$.
The factors $1/m_D$ and $1/m_D^3$ are included 
to ensure appropriate normalizations 
for the wave functions 
(dimensionless overlap integral functions).

The previous radial wave functions are normalized 
according with~\cite{NDeltaD,LatticeD} 
\ba
& &
\int_k | \psi_S (\bar P,k) |^2 = 1, \nonumber \\
& &
\int_k \tilde k^4 | \psi_{D3} (\bar P,k) |^2 =1, 
\label{eqNorma}\\
& &
\int_k \tilde k^4 | \psi_{D1} (\bar P,k) |^2 =1,
\nonumber
\ea
where $\bar P$ represents the $\Omega^-$ momentum 
at the rest frame: $\bar P = (M_\Omega,0,0,0)$ 
and $\tilde k =(0, {\bf k})$.

The inspiration for Eq.~(\ref{eq-psiS}) comes from 
the representation of the nucleon radial wave function,
since it can be reduced to the Hulthen
form in the non relativistic limit 
in the configuration space (difference of two Yukawa functions)~\cite{Nucleon}.
Compared to our previous work on the $\Omega^-$ baryon
with $D$ states~\cite{Omega2}, we kept the expression for 
the $D3$ and $D1$ radial wave functions but modified the form for $\psi_S$,
which is now defined by the product of two different multipoles on 
the variable $\chi$.

The motivation to the new form is twofold:
generate an asymptotic form for the $S$-state contribution 
compatible with the leading-order form factors
at large $Q^2$ (proportional to $1/Q^4$), 
and increase the flexibility of the fit,
including two different momentum scale parameters.
The parametrizations (\ref{eq-psiS})--(\ref{eq-psiD1}) 
are compatible with the expected falloffs 
(\ref{eqGE0lQ2}) and  (\ref{eqGE2lQ2}) 
of the form factors for large $Q^2$, as discussed below.
In the previous work~\cite{Omega2}, our main goal 
was the description  of the $\Omega^-$ 
in a limited region of $Q^2$.

The parametrizations (\ref{eq-psiS})--(\ref{eq-psiD1}), may, however, 
be incompatible with the relation (\ref{eqGM3extra}).
For that reason, we consider also 
an alternative parametrization of the $S$-state wave function,
and minor modifications in the $D$-state radial wave functions.

\subsection{Alternative parametrizations for the radial wave functions}

Alternative parametrizations for the 
$S$, $D3$, and $D1$ radial wave functions can be
\ba
& & 
\hspace{-.05cm}
\psi_{S} (P,k) = 
\frac{N_S}{m_D}
\left[
\frac{1}{(\alpha_1^\prime + \chi) (\alpha_2^\prime + \chi)^2} 
- \frac{r_S}{(\alpha_1 + \chi)(\alpha_2 + \chi)}
\right], \nonumber \\
&& \label{eq-psiSb}\\
& & 
\hspace{-.05cm}
\psi_{D3} (P,k) = 
\frac{N_{D3}}{m_D^3}
\frac{1}{(\alpha_1 + \chi)(\alpha_2 + \chi)(\alpha_3 + \chi)^2}, 
\label{eq-psiD3b} \\
& & 
\hspace{-.05cm}
\psi_{D1} (P,k) = 
\frac{N_{D1}}{m_D^3}
\frac{1}{(\alpha_1 + \chi)(\alpha_2 + \chi)(\alpha_4 + \chi)^2}, 
\label{eq-psiD1b}
\ea
where $N_S$, $N_{D3}$, and $N_{D1}$ are new normalization constants,
$\alpha_1^\prime$ and $\alpha_2^\prime$ 
are additional square momentum range parameters, and  
$r_S$ is a new adjustable parameter.
The normalization constants are also determined by 
the conditions (\ref{eqNorma}).

The parametrizations (\ref{eq-psiSb})--(\ref{eq-psiD1b}) 
are characterized by the common factors 
associated with the parameters $\alpha_1$ and  $\alpha_2$.
In Sec.~\ref{sec-Results-lQ2}, we show that in Eq.~(\ref{eq-psiSb}), 
the term associated with the parameters
$\alpha_1$ and $\alpha_2$ dominates over the 
term with the associated with parameters 
$\alpha_1^\prime$ and $\alpha_2^\prime$, 
on the overlap integral, at large $Q^2$.

\subsection{$\Omega^-$ form factors}
\label{secFF}

The $\Omega^-$ elastic form factors are calculated 
in a previous work~\cite{Omega2} for 
a mixture of $S$, $D3$, and $D1$ states given by Eq.~(\ref{eqPsiTotal}),
in the first order of the coefficients $a$ and $b$.
The approximation is justified for small 
admixture coefficients.

To represent the $\Omega^-$ elastic form factors,
it is convenient to define 
the functions~\cite{DeltaSFF,DeltaFF,Omega2}
\ba
\tilde e_{\Omega} = -f_{10}(Q^2),
\hspace{.5cm}
\tilde \kappa_{\Omega} = - \frac{M_\Omega}{M_N}f_{20}(Q^2).
\ea
We use the tilde to represent functions of $Q^2$
without the explicit inclusion of the argument.
Also useful for the representation of the 
electric-type and magnetic-type form factors 
are the combinations
\ba
& &
\tilde g_\Omega = \tilde e_\Omega - \tau \tilde \kappa_\Omega, 
\label{eq-gtilde}\\
& &
\tilde f_\Omega = \tilde e_\Omega + \tilde \kappa_\Omega,
\label{eq-ftilde}
\ea
where $\tau= \frac{Q^2}{4M_\Omega^2}$.
Note the similarity with the expressions for 
the electric and magnetic form factors of  
$\frac{1}{2}^+$ baryons.

The covariant spectator quark model 
results for the $\Omega^-$ elastic form factors
can now be written as~\cite{Omega2}
\ba
G_{E0}(Q^2) &=& N^2 \, \tilde g_\Omega \,  {\cal I}_S, 
\label{eqGE0}\\
G_{M1}(Q^2) &=&  N^2 \, \tilde f_\Omega \left[ {\cal I}_S  +
\frac{4}{5} a \, {\cal I}_{D3} - \frac{2}{5} b \, {\cal I}_{D1} 
\right], 
\label{eqGM1}\\
G_{E2} (Q^2)&=& N^2 \, \tilde g_\Omega \,  (3a)  \frac{{\cal I}_{D3}}{\tau},
\label{eqGE2}\\
G_{M3} (Q^2)&=& 
   N^2 \, \tilde f_\Omega \left[ 
a  \frac{{\cal I}_{D3}}{\tau} + 2b \frac{{\cal I}_{D1}}{\tau}
\right],
\label{eqGM3}
\ea
where $N^2$ is a normalization factor, and 
the overlap integral functions are determined by
\ba
& &
{\cal I}_S = \int_k \psi_S(P_+,k)  \psi_S(P_-,k), 
\label{eqIntS} \\
& &
{\cal I}_{D3} = \int_k b(\tilde k_+,\tilde q_+)
\psi_{D3}(P_+,k)  \psi_S(P_-,k), 
\label{eqIntD3} \\
& &
{\cal I}_{D1} = \int_k b(\tilde k_+,\tilde q_+)
\psi_{D1}(P_+,k)  \psi_S(P_-,k).
\label{eqIntD1} 
\ea
The function $b(\tilde k_+,\tilde q_+)$ has the form 
\ba
b(\tilde k_+,\tilde q_+) = 
\frac{3}{2} \frac{(\tilde k_+ \cdot \tilde q_+)}{\tilde q_+^2}
- \tilde k_+^2,
\ea
where $\tilde k_+ = k - \frac{P_+ \cdot k}{M_\Omega^2} P_+$
and 
$\tilde q_+ = q - \frac{P_+ \cdot q}{M_\Omega^2} P_+$~\cite{DeltaFF}.

For future discussion, it is worth mentioning  
that the function $b(\tilde k_+,\tilde q_+)$ 
reduces to $- \sfrac{1}{2}{\bf k}^2 (1- 3 z^2)$,
in the final-state rest frame, where $z$ is the cosine 
of the angle between ${\bf k}$ and ${\bf q}$.
We recover then the dependence on the spherical 
harmonic $Y_{20}(z) \propto (3 z^2 - 1)$ as in 
the non relativistic limit.

From (\ref{eqGE0})--(\ref{eqGM3}), we can conclude 
that the state $D3$ is responsible by the nonzero results 
of the electric quadrupole form factor
and that both $D3$ and $D1$ contribute to the 
magnetic octupole form factor.
In the limit $a=0$ (no $D3$ state) and $b=0$ 
(no $D1$ state), 
we recover the results of an $S$-state model
with $G_{E2} \equiv 0$ and $G_{M3} \equiv 0$~\cite{DeltaSFF,Omega,Deformation}.

In the expressions for the electromagnetic form factors 
(\ref{eqGE0})--(\ref{eqGM3}), notice that the functions $G_{E2}$ and $G_{M3}$ 
include dependence on ${\cal I}_{D3}/\tau$ and ${\cal I}_{D1}/\tau$.
These form factors are, however, well defined 
in the limit $\tau \to 0$ ($Q^2 \to 0$),
since we can show that  
${\cal I}_{D3} \propto Q^2$ and ${\cal I}_{D1} \propto Q^2$ 
near $Q^2=0$~\cite{DeltaFF}.

We can now discuss the normalization factor $N^2$.
When the baryon elastic form factors are calculated 
in the first order of $a$ and $b$, and drop terms 
of the order of $a^2$ and $b^2$, one should take $N^2=1$.
We notice, however, that in the limit $Q^2=0$,
we can calculate the baryon charge using all orders of $a$ and $b$.
In that case, we obtain $N^2=1/(1 + a^2 + b^2)$.
In the previous work, 
we choose to perform the calculations with $N^2=1/(1 + a^2 + b^2)$,
and include a theoretical band where the upper limit 
is determined by $N^2=1$.
With this procedure, one obtains consistent results for 
$G_{E0}(0)$ which must reproduce the $\Omega^-$ 
electric charge [${\cal I}_S \to 1$, 
$G_{E0} (0) \to  \tilde e_\Omega \to -1$].

In the present work, we simplify the previous procedure, 
assuming that the best estimate is the average
between $N^2= 1$ and $N^2=1/(1 + a^2 + b^2)$.
Our central value is then determined by
\ba
N^2 = \frac{1 + \frac{a^2}{2}  +  \frac{b^2}{2}}{1 + a^2 + b^2}.
\ea
To take into account the  theoretical uncertainty, 
we use $N^2=1$ to the upper limit, as before,  
and \mbox{$N^2=1/(1 + a^2 + b^2)$} for the lower limit.
The new procedure favors the fit to the data, 
since the fit to the $G_{E0}$ lattice data at low $Q^2$ 
is improved when $N^2$ is \mbox{closer to 1.}

\subsection{Large $Q^2$ behavior}
\label{secLargeQ2}

One can now look for the asymptotic form of the 
$\Omega^-$ electromagnetic form factors.
The analysis of the overlap integrals 
${\cal I}_S$, ${\cal I}_{D3}$ and  ${\cal I}_{D1}$,
shows that
\ba
{\cal I}_S, \;  {\cal I}_{D3}, \; 
{\cal I}_{D1} \propto \frac{1}{Q^4},
\label{eqInt-infty}
\ea
apart logarithmic corrections.
These asymptotic falloffs are valid for the 
parametrizations (\ref{eq-psiS})--(\ref{eq-psiD1}) 
and (\ref{eq-psiSb})--(\ref{eq-psiD1b}).

Taking the previous results into account,
one can conclude from Eqs.~(\ref{eqGE0})--(\ref{eqGM3})
that the form factors are ruled by Eqs.~(\ref{eqGE0lQ2}) and (\ref{eqGE2lQ2}) 
for large $Q^2$: $G_{E0}, G_{M1} \propto 1/Q^4$ 
and $G_{E2}, G_{M3} \propto 1/Q^6$, 
apart logarithmic corrections.

The results (\ref{eqInt-infty}) are the consequence of the combination 
of the form of the radial wave functions in the overlap integrals 
and the fact that in the first order in the 
coefficients $a$ and $b$ all overlap integrals have 
at least a contribution of the  $S$-state radial wave function.
Since the overlap integrals are invariant, 
the integrals can be performed in any frame.
The calculations are simplified when 
we choose the frame where the $S$-state is at rest.
In these conditions, one can prove that 
if we use an $S$ state radial wave functions 
with the form $\psi_S \propto 1/(\alpha + \chi)^3$
one obtains for the overlap integrals,
falloffs with $1/Q^6$, without logarithmic corrections.
This result is derived in Appendix G from Ref.~\cite{NDelta}.
If we consider instead $\psi_S \propto 1/(\alpha + \chi)^2$, 
one concludes that the overlap integrals are  
dominated by terms  on $\frac{1}{Q^4}\left(\log \frac{Q^2}{M_B^2} \right)$,
where $M_B$ is the mass of the baryon at rest
(also demonstrated in Appendix G from Ref.~\cite{NDelta}).
In a present case, where the $S$-state 
radial wave function has a term  
$\psi_S \propto 1/(( \alpha_1 + \chi)(\alpha_2 + \chi))$,
we can conclude that the overlap integrals are 
also dominated at large $Q^2$ with terms of the order 
$1/Q^4$ with logarithmic corrections~\cite{Note1}.

The possibility of verification of the condition 
(\ref{eqGM3extra})
is discussed in Sec.~\ref{sec-Results-lQ2}.
In that section, we analyze also the results 
obtained when we use 
the radial wave functions (\ref{eq-psiSb})--(\ref{eq-psiD1b}),
and discuss the motivation to those expressions.

\begin{table*}[t]
\begin{tabular}{l  c  c c c c c}
\hline
\hline
    & $a$ & $b$ &
$\alpha_1$ & $\alpha_2$ & $\alpha_3$ & $\alpha_4$      \\
\hline
\hline
Fit SL data             & 0.0322 & 0.2776 &  0.05927 & 0.1075   &  0.4437  & 0.5375  \\     
Fit SL/TL data          & 0.0304  & 0.2307 & 0.04250 &  0.1482  &  0.3340 &  0.2485  \\  
\hline
\hline
\end{tabular}
\caption{\footnotesize 
Adjustable parameters of the fits of the spacelike (SL) and timelike (TL) data.
Included are the admixture coefficients $a$ and $b$ of the $D3$ and  $D1$ states,
and the parameters associated with the  $S$- ($\alpha_1$, $\alpha_2$),
$D3$- ($\alpha_3$) and $D1$-state ($\alpha_4$) radial wave functions,
defined by Eqs.~(\ref{eq-psiS})--(\ref{eq-psiD1}).}
\label{tableParam}
\end{table*}

\begin{table*}[t]
\begin{tabular}{l  c  c c c c c c c c}
\hline
\hline
    & $G_{E0}$ & $G_{M1}$ &  $G_{E2}$ & $G_{M3}$ & $|G|$  & Total & $\%D1$ &    $\%D3$ \\
\hline
\hline
Fit SL data              & 2.64 & 1.24 &  0.25  & 2.37  &  &  {\bf 1.54} & 0.096 & 7.15  \\     
Fit SL/TL data           & 2.53 & 1.66 &  0.37  & 3.40  &  2.79 & {\bf 1.74} & 0.088 & 5.05 \\  
Fit SL/TL + LQ2          & 2.54 & 1.92 &  0.48  & 5.33  &  3.33 & {\bf 1.91} & 0.092 & 7.55 \\  
\hline
\hline
\end{tabular}
\caption{\footnotesize The quality of the fit 
measured by the chi square per data point  and percentages 
of $D1$ and $D3$ states. $|G|$ represents the timelike data.
LQ2 labels the large-$Q^2$ condition (\ref{eqGM3extra}).}
\label{tableChi2}
\end{table*}

\section{Electromagnetic form factors in the 
timelike and spacelike regions}
\label{secOmega-TL}

In the present section, we tested if the parametrizations 
discussed in the previous section are suitable to describe 
the available $\Omega^-$ electromagnetic form factor data.

We divide the study in three steps:
\begin{itemize}
\item
First, we tested if the simplest 
parametrizations, 
based on radial wave functions (\ref{eq-psiS})--(\ref{eq-psiD1}),
defined by the asymptotic forms $G_{E0}$, $G_{M1} \propto 1/Q^4$ 
and $G_{E2}$, $G_{M3} \propto 1/Q^6$ 
are compatible with the lattice QCD data 
[spacelike (SL) region].
\item
In a second step, we tested if the same kind 
of parametrization can also describe the 
timelike data,
more specifically the data associated with 
the effective form factor $|G (q^2)|$ from CLEO
[timelike (TL) region].
\item  
At the end, we tested if a modified model 
based on the  radial wave functions  (\ref{eq-psiSb})--(\ref{eq-psiD1b})
is consistent with the large-$Q^2$ condition (\ref{eqGM3extra})
and debate the range of $Q^2$ where the relation 
can be fulfilled.
[In the following, we use LQ2 to label 
the large-$Q^2$ condition (\ref{eqGM3extra})].
\end{itemize}

In a previous study~\cite{Omega2}, we considered only 
the lattice data in the region $Q^2 \le 1$ GeV$^2$
for $G_{E0}$ and $G_{M1}$, from Ref.~\cite{Alexandrou10},
because we were more focused in the 
\mbox{low-$Q^2$} behavior of the form factors, 
including the results for $G_{E2}(0)$ and $G_{M3} (0)$.  
In the present work, we extend the range to $Q^2 \le 2$ GeV$^2$,
in order to take into account large-$Q^2$ effects on 
the form factors, which are pertinent to the timelike region.
We discard the lattice QCD data for $Q^2 > 2$ GeV$^2$,
because those simulations are affected by very large error bars, 
and cannot be used to discriminate between different parametrizations.

Since Ref.~\cite{Alexandrou10} presents no data for $G_{M3}$,
we include in our database the $G_{M3}$ data point from 
Boinepalli \etal~\cite{Boinepalli09} for $Q^2= 0.23$ GeV$^2$,
even though the result is not very accurate.

\subsection{Adjust parameters to the spacelike data}

We adjust the free parameters of the model,
the admixture coefficients $a$, $b$ and 
the parameters of the radial wave functions (\ref{eq-psiS})--(\ref{eq-psiD1}),
$\alpha_i$  ($i=1,..,4$), to the spacelike data from our database.
The value associated with the magnetic moment (\ref{eqGM10})
is not included in the present fit because 
it was already used in the calibration
of the strange quark current (fixes $\kappa_s$)~\cite{Omega}.
The parameters of the best fit are presented in 
the first row of Table~\ref{tableParam} (Fit SL data).

The quality of the fit, estimated by the chi square 
per data point for the different subsets of data 
($G_{E0}$, $G_{M1}$, $G_{E2}$, and $G_{M3}$), is presented 
in the first row of Table~\ref{tableChi2}.
In the column ``Total'', we present the total chi square per data point.

Compared with our previous study of the $\Omega^-$
form factors from Ref.~\cite{Omega2},
we obtain a better description of the 
$G_{E0}$, $G_{M1}$ form factors and improve 
the overall description of the data 
(smaller total chi square per data point).
There are three main reasons for this improvement:
because we increase the range of the $Q^2$ lattice data,
and the chi square associated to the large $Q^2$ is smaller,
because we consider a radial wave function for $\psi_S$
(\ref{eq-psiS}) with a falloff which better describe the data,
and also because $\psi_S$ 
include two momentum range scales (an extra parameter).
The most relevant factor to this improvement  
is the form of radial wave function,
since we replaced a tripole form $1/(\alpha_i + \chi)^3$ 
in Ref.~\cite{Omega2} by a product of two monopoles $1/(\alpha_i + \chi)$.
This conclusion was confirmed by numerical calculations.

\begin{figure*}[t]
\vspace{.5cm}
\centerline{
\mbox{
\includegraphics[width=3.2in]{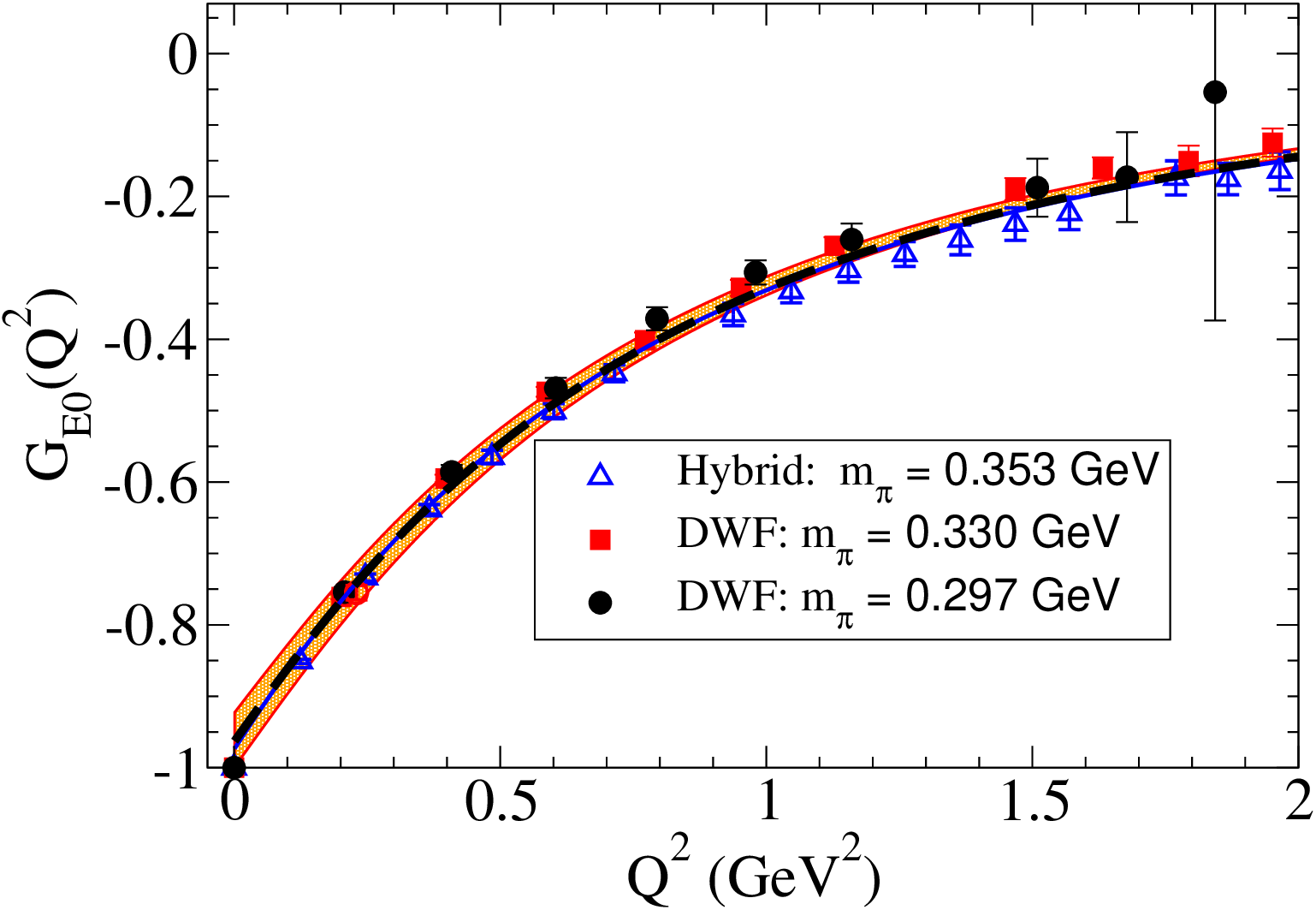} \hspace{.7cm}
\includegraphics[width=3.15in]{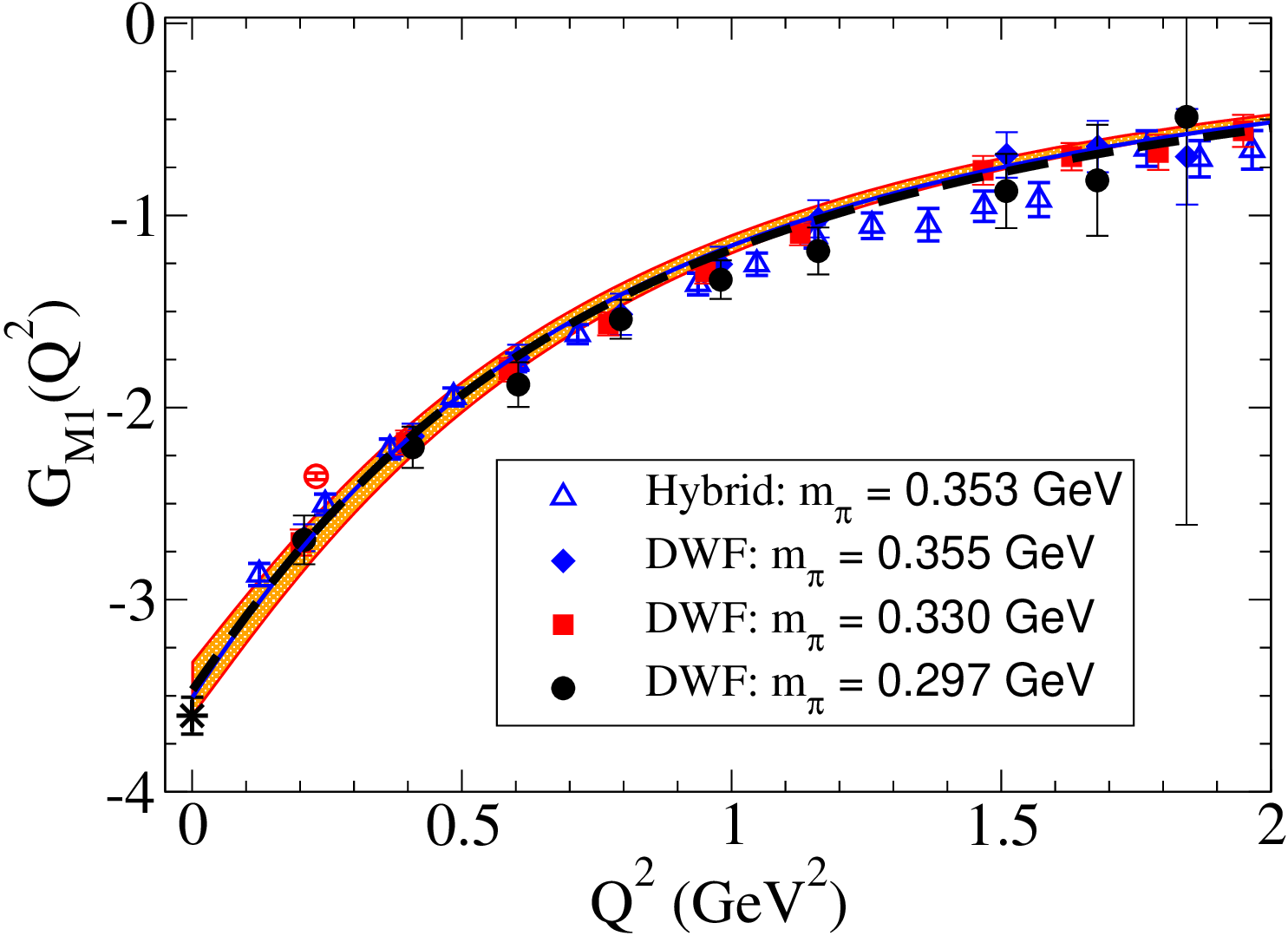}
}}
\caption{\footnotesize{Form factors $G_{E0}$ and $G_{M1}$.
Fits to the data: SL (dashed line), SL/TL (solid line)
and SL/TL + LQ2 (orange band).
Lattice QCD data from Alexandrou \etal~\cite{Alexandrou10}.
For $G_{M1}$ we include also the experimental result 
$G_{M1}(0)= -3.60   \pm 0.09$~\cite{PDG20} ($\ast$).
The open circles represent the 
result for $Q^2=0.23$ GeV$^2$ from
Boinepalli \etal~\cite{Boinepalli09}.}}
\label{figFF1}
\end{figure*} 
\begin{figure*}[t]
\vspace{.5cm}
\centerline{
\mbox{
\includegraphics[width=3.2in]{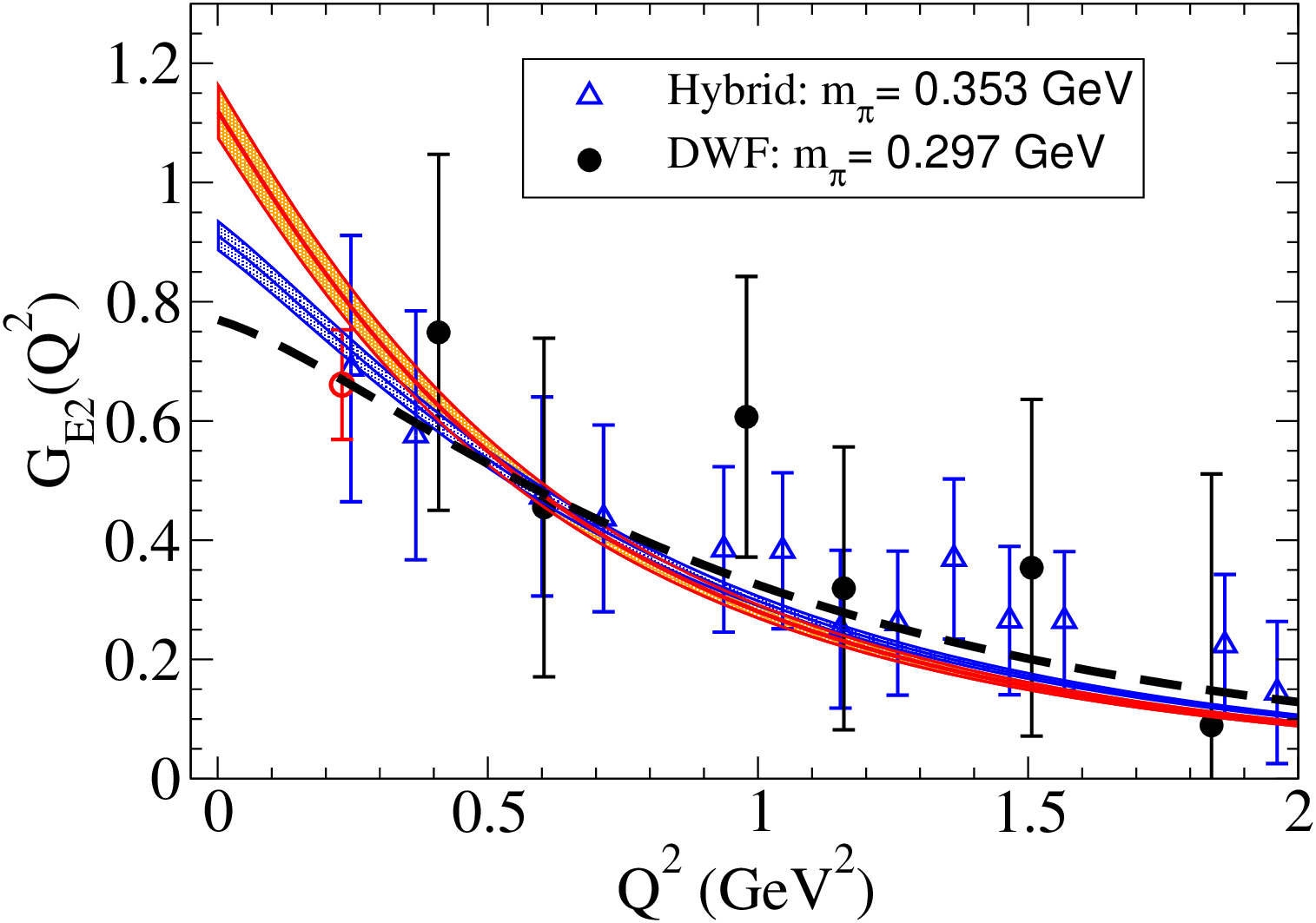} \hspace{.7cm}
\includegraphics[width=3.15in]{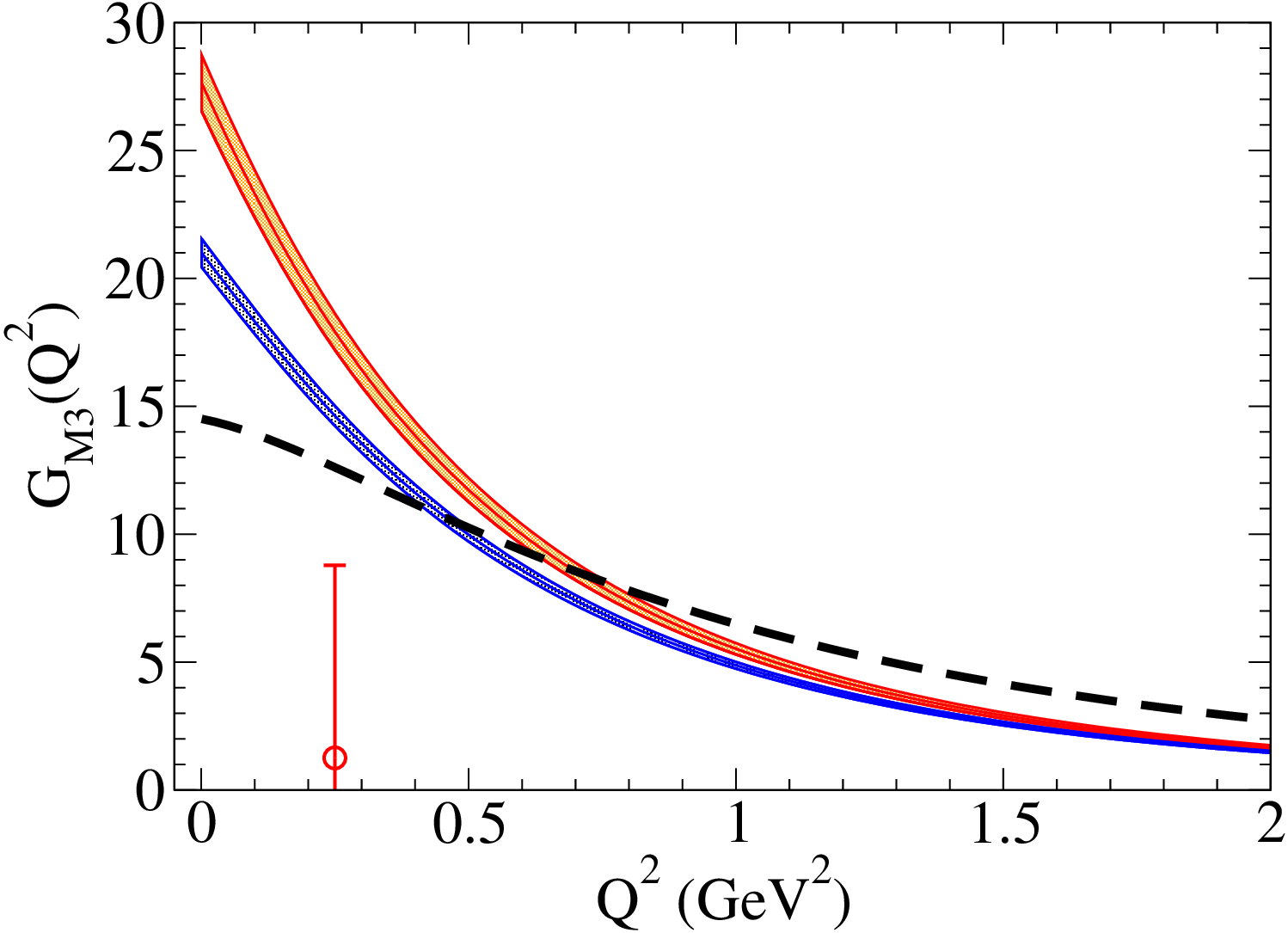}
}}
\caption{\footnotesize{Form factors $G_{E2}$ and $G_{M3}$
Fits to the data: SL (dashed line), 
SL/TL (blue solid line/blue band) and SL/TL + LQ2 (red solid line/orange band). 
Lattice QCD data from Alexandrou \etal~\cite{Alexandrou10}.
The open circles represent the 
result for $Q^2=0.23$ GeV$^2$ from
Boinepalli \etal \cite{Boinepalli09}.}}
\label{figFF2}
\end{figure*}

The improvement in the description of 
the data with the new parametrization for $\psi_S$ is pertinent 
because it shows that a radial wave function compatible 
with the large-$Q^2$ pQCD behavior, Eqs.~(\ref{eqGE0lQ2}) and (\ref{eqGE2lQ2}),
can improve also the description of the low-$Q^2$ region.
We emphasize that the use of two scales, 
$\alpha_1$, $\alpha_2$ in the radial wave function,
instead of one global scale, $\alpha_1$, as in Ref.~\cite{Omega2}, 
also contributes to the improvement.

In Table~\ref{tableChi2}, we include also the 
relative contributions from the $D1$ and $D3$ states.
We notice that the values of $a$ and $b$ 
are very close to the values of the previous work: 
$a=0.0341$ and $b=0.2666$~\cite{Omega2} (see Table~\ref{tableParam}).
It is then worth mentioning that, although based 
on different $S$-state radial wave functions,  
the mixtures of the $D1$ and $D3$ states, 
the fit from Ref.~\cite{Omega2}, and the new fit are very similar,
with about 0.1\%  and 7\% for the $D3$ and $D1$ states, respectively.
The conclusion that the $D1$ state has a larger contribution is preserved. 
Notice, however, that the  0.1\% of the state $D3$ 
is essential to describe the electric quadrupole form 
factor data, according to Eq.~(\ref{eqGE2}).

We look now for the numerical results for 
the form factors for $G_{E0}$, $G_{M1}$, $G_{E2}$ and  $G_{M3}$, 
represented in Figs.~\ref{figFF1} and \ref{figFF2} by the dashed lines.
We do not include the theoretical uncertainty band for clarity.
As anticipated from the results for the chi squares, 
we obtain a good description of the $G_{M1}$ and $G_{E2}$ data.
The $G_{E0}$ lattice data are more difficult to describe, 
due to the behavior of the different datasets and the small error bars.

The comparison with the $G_{E2}$ and $G_{M3}$ lattice QCD data
are presented in  Fig.~\ref{figFF2}.
The fit (dashed-line) describes well the 
lattice QCD data for $G_{E2}$ within the 
accuracy of the data points.
The small value obtained for the $G_{E2}$ partial chi square (0.25)
is the consequence of the large error bars on the data.
As for $G_{M3}$, the estimate can be compared 
only with the single data point 
$G_{M3}(\, 0.23 \; \mbox{GeV}^2)= 1.25 \pm 7.50$~\cite{Boinepalli09}.
The present result overestimates the data,
but only by 1.1 standard deviations.

When compared with the estimate from Ref.~\cite{Omega2}
one has an increment of 13\% for $G_{E2}(0)$ 
and a reduction of 6\% on  $G_{M3}(0)$.
The falloffs of $G_{E0}$ and $G_{M3}$ with $Q^2$
are similar to the ones from Ref.~\cite{Omega2}.

\subsection{Adjust parameters to the spacelike and timelike data}
\label{secSLTLfit}

In the previous section, we demonstrated
that the covariant spectator quark model
is successful in the description of the $\Omega^-$ spacelike data.

One can notice, however, that the test of the model 
is restricted to the range $0 \le Q^2 \le 2$ GeV$^2$.
The model was not tested in the large-$Q^2$ region,
because the lattice QCD simulations are limited in the range of $Q^2$, 
and it was also not tested in the timelike region.

The next step is to test if the derived parametrization 
is consistent with the $\Omega^-$ timelike data 
obtained from $e^+ e^- \to \Omega^- \bar \Omega^+$ 
cross section data in CLOE~\cite{Dobbs17a},
expressed in terms of the effective form factor $|G(q^2)|$. 
This test was performed with the model parametrization 
from Ref.~\cite{Omega2} in Ref.~\cite{Hyperons}.
The conclusion was that either the value of $G_{M3}(0)$
is overestimated or the form factors drop off 
much faster that in the original parametrization.

The new timelike data provide then, a unique opportunity
to study the magnitude of $G_{M3}(0)$ and 
the falloff of the $\Omega^-$ form factors,
which cannot be tested by the available lattice QCD
(limited in the range of $Q^2$).

To extend the calculations of the covariant spectator 
quark model to the timelike region 
($q^2= -Q^2 > 0$), we use the asymptotic relations 
proposed in Ref.~\cite{Hyperons}
for the electric ($\ell =0,2$)
and magnetic ($\ell = 1,3$) form factors 
\ba
G_{E\ell}^{\rm TL} (q^2) &=& G_{E\ell} (Q^2 + 2 M_\Omega^2 ), 
\label{eqGEass}
\\
G_{M\ell}^{\rm TL} (q^2) &=& G_{M\ell} (Q^2 + 2 M_\Omega^2),
\label{eqGMass}
\ea
where the index TL indicates the timelike form factors.
On the rhs $G_{E \ell}$ and $G_{M \ell}$
represent the spacelike form factors.
To calculate the effective form factors $|G (q^2)|$ 
of $\frac{3}{2}^+$ baryons, 
we use the replacements 
(\ref{eqGE2abs}) and  (\ref{eqGM2abs}),
and the $G_{E \ell}$ and $G_{M \ell}$ spacelike form factors~\cite{Korner77}.
The relations (\ref{eqGE2abs}) and  (\ref{eqGM2abs})
are derived from general physics 
and mathematical principles, 
including unitarity and the Phragm\'en-Lindel\"{o}f theorem,
valid for analytic function of $q^2$ for 
very large $|q^2|$~\cite{Pacetti15a,Denig13}.
A consequence of the approximation 
is that the form factors are also real functions 
in the timelike region for large $q^2$.
The shifts of $2 M_\Omega^2$ on the rhs of Eqs.~(\ref{eqGE2abs}) 
and (\ref{eqGM2abs}) are motivated 
by the difference between the spacelike ($Q^2=0$)
and timelike thresholds ($q^2= 4 M_\Omega^2$).
The relations used here include then finite $q^2$ 
corrections to the asymptotic limit~\cite{Hyperons}.

The parameters associated with the global fit 
are presented in the last row of Table~\ref{tableParam}.
The corresponding chi square per data point 
for each form factor, is included 
in the second row of Table~\ref{tableChi2} (Fit SL/TL data).
The column  $|G|$ indicates the chi square
associated with the timelike data.

An interesting result from Table~\ref{tableChi2} 
is that the consideration of the timelike data leads to the improvement
of the description of the $G_{E0}$ data,
meaning that the falloff of $G_{E0}$ is relevant 
for the description of the timelike data.
The differences between the two parametrizations
(SL or SL/TL) are related to the variation in about 30\% 
on $\alpha_1$, as one can see in Table~\ref{tableParam}.
Recall that $G_{E0}$ depend exclusively on the $S$ state.
These differences, however, are not perceived on the graph for $G_{E0}$ 
below $Q^2=2$ GeV$^2$.

The comparison of the global fit (SL/TL) with the 
spacelike data is presented in Figs.~\ref{figFF1} and \ref{figFF2}.
In Fig.~\ref{figFF1}, we include only the blue solid line for clarity.
The theoretical uncertainty 
associated with the blue line
has a magnitude similar to the orange band, 
discussed in the next subsection.
In Fig.~\ref{figFF2}, we present the result of the fit by 
the blue band in order to include the theoretical error 
associated with the normalization factor $N^2$, 
as discussed in Sec.~\ref{secFF}.

The results of the fit to the SL and TL data for $G_{E0}$ 
and $G_{M1}$ (solid line) are very similar to the results 
of the fit to the SL data (dashed line).
We conclude then that the main differences 
between the two fits (SL and SL/TL) appear only 
for values of $Q^2$, larger than 2 GeV$^2$.

The results for $G_{E2}$ and $G_{M3}$ displayed in 
Fig.~\ref{figFF2} are more relevant.
One can notice the increasing of $G_{E2}$ and $G_{M3}$ 
near $Q^2=0$ and a stronger falloff with $Q^2$
in comparison with the fit which ignore the timelike data.
The quality of the description of the $G_{E2}$ data 
are not substantially modified (large error bars, partial 
chi square per data point of 0.2--0.4).
As for $G_{M3}$ the impact of the timelike data is more significant.
Notice that the quality of the description of the 
$G_{M3}$ data point is reduced in comparison with 
the previous fit (dashed line).
The estimate differs from the data by 1.8 standard deviations.
When we take into account the timelike data, 
we deteriorate the description of the $G_{M3}$ data, represented by 
the point with $Q^2=0.23$ GeV$^2$.

\begin{figure}[t]
\vspace{.5cm}
\centerline{
\mbox{\includegraphics[width=3.2in]{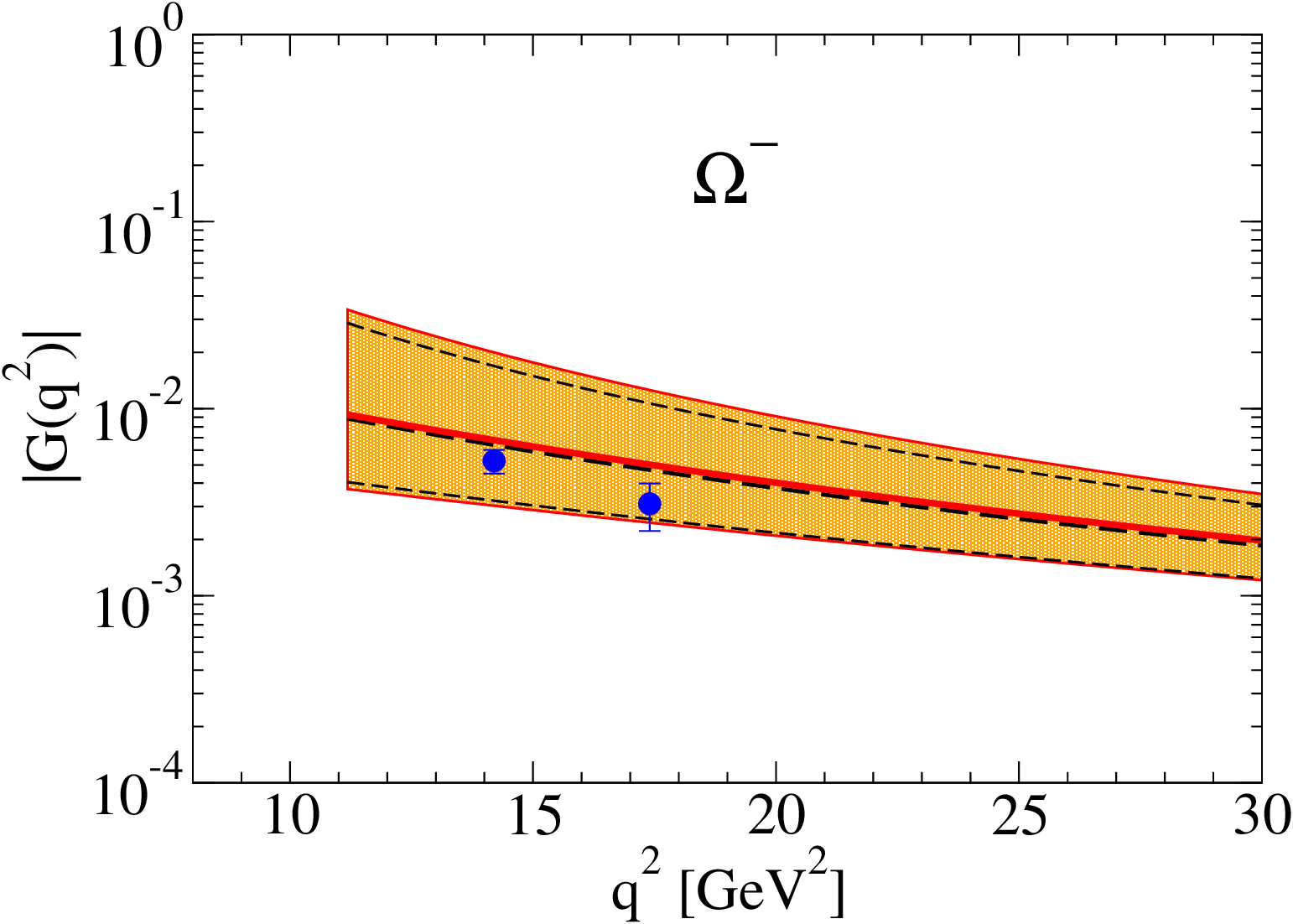} }}
\caption{\footnotesize{Results 
for the $\Omega^-$ effective form factor $|G(q^2)|$ 
in the timelike region ($q^2 > 0$).
The thick dashed line indicates the result of
the fit SL/TL which ignore the large $Q^2$ constraints,
and the thin dashed lines represent the 
upper and lower limits (theoretical error). 
The solid line is the best estimate (fit SL/TL + LQ2) based 
on asymptotic relations, including the theoretical error (orange band).
The data are from CLEO~\cite{Dobbs17a}.
}}
\label{figGT-Omega}
\end{figure}

The main difference in the new solution to $G_{M3}$ 
is in the faster falloff with $Q^2$, indicating that
the magnitude of the function is significantly 
reduced for large $Q^2$, contributing to a significant suppression 
of the term $|G_M|^2$ in $|G(q^2)|$ 
[see Eq.~(\ref{eqGM2abs})].

The results for the effective form factor $|G (q^2)|$ 
for the $\Omega^-$ are presented in Fig.~\ref{figGT-Omega},
and compared with the more recent data from CLOE
for $q^2= 14.2$ and 17.4 GeV$^2$~\cite{Dobbs17a}.
We omitted the first measurement at $q^2= 14.2$ GeV$^2$ 
from CLOE~\cite{Dobbs14a}, since it is superseded
by a more recent analysis~\cite{Dobbs17a}. 
The dashed line corresponds to our best estimate 
based on Eqs.~(\ref{eqGEass}) and (\ref{eqGMass})
and the spacelike and timelike data.
The thin dashed lines indicate the theoretical error 
based on finite correction on $q^2$ for the asymptotic relations
between spacelike and timelike regimes,  
which are valid strictly when $q^2 \to \pm \infty$. 
We estimate these errors using the replacements 
on Eqs.~(\ref{eqGEass}) and (\ref{eqGMass}):
 $Q^2 + 2 M_\Omega^2 \to Q^2$ for the upper limit
and $Q^2 + 2 M_\Omega^2 \to Q^2 + 4 M_\Omega^2$ 
for the lower limit~\cite{Hyperons}.
At very large $Q^2$, the width of the variation
become negligible and the 
leading-order falloff of $G$, $1/q^4$, emerges~\cite{Hyperons}.

We return to this discussion after taking into account the impact of the 
large-$Q^2$ relation (\ref{eqGM3extra}) on the structure of the form factors.

\subsection{Model compatible with the asymptotic conditions}
\label{sec-Results-lQ2}

In Sec.~\ref{secLargeQ2}, we conclude that the 
leading-order dependence of the electromagnetic form factors 
at large $Q^2$: $G_{E0}$, $G_{M1} \propto 1/Q^4$ and $G_{E2}$, 
$G_{M3} \propto 1/Q^6$, is naturally reproduced 
with the radial wave functions (\ref{eq-psiS})--(\ref{eq-psiD1}).
One can then ask if the large $Q^2$ constraint (\ref{eqGM3extra}) 
can also be accomplished with the same model 
for the radial wave functions.
The answer is no, as demonstrated below.

For the following discussion, it is necessary to know 
that  $\tilde f_\Omega$ is a negative function of $Q^2$
(see Appendix~\ref{appModel}) 
and the signs of the overlap integrals 
${\cal I}_S$ and ${\cal I}_{D1}$ defined 
by (\ref{eqIntS}) and (\ref{eqIntD1}).
Based on the parametrizations discussed 
in the previous subsections, we assume 
also that the mixture coefficients $a$ and $b$ are positive,
as suggested by the previous analysis (SL and ST/TL fits).

The integral ${\cal I}_S$ is positive by construction
(positive integrate functions), and consequently it does not change the sign.
The integrals ${\cal I}_{D3}$ and  ${\cal I}_{D1}$ 
include the angular function \mbox{$Y_{20}(z) \propto  (3 z^2 -1)$,}
when we consider the final-state rest frame.
In the limit $Q^2=0$ 
the integrals ${\cal I}_{D3}$ and  ${\cal I}_{D1}$ 
vanish due to the factor $Y_{20}(z)$.
For finite $Q^2$, the $D$-state integrals 
can be positive or negative 
depending on $Q^2$ and on the square momentum 
parameters of the radial wave functions ($\alpha_i$, $i=1,...,4$).
For the typical values of the parameters 
(determined from the fits to the $0 \le Q^2 \le 2$ GeV$^2$ region),
one concludes that the functions ${\cal I}_{D1}$,
as ${\cal I}_{D3}$, do not change sign in wide a range of $Q^2$.
Numerical calculations suggest that the possible 
zeros for ${\cal I}_{D3}$ and  ${\cal I}_{D1}$ 
appear only for $Q^2 > 10^6$ GeV$^2$,
well above the present-day range of experiments.
Thus, for the purpose of the applications 
of the parametrizations (\ref{eq-psiS})--(\ref{eq-psiD1}), 
we can assume that ${\cal I}_{D3}$ and  ${\cal I}_{D1}$ 
are functions with a defined sign
(do not change sign). 

We can now explain why the parametrizations 
(\ref{eq-psiS})--(\ref{eq-psiD1})
are incompatible with the relation (\ref{eqGM3extra}).
When we combine Eq.~(\ref{eqGM1}) and Eq.~(\ref{eqGM3})
with the condition (\ref{eqGM3extra}), 
we conclude that the last condition 
is valid if ${\cal I}_S = 2 b \, {\cal I}_{D1}$
(see Appendix~\ref{appModel}).
A consequence of the previous relation is that
the contribution from the $D3$ state 
is not relevant for Eq.~(\ref{eqGM3extra}).

The verification of Eq.~(\ref{eqGM3extra}) implies that 
$G_{M1}$ and $G_{M3}$ must have the same sign for large $Q^2$. 
The $S$-state gives the dominant positive contribution to $G_{M1} < 0$, 
while the state $D1$ gives a large contribution to $G_{M3} > 0$, near $Q^2=0$,
implying that $b \, {\cal I}_{D1} < 0$, since $\tilde f_\Omega < 0$.
The conclusion is then that ${\cal I}_S >0$ 
and  $b \, {\cal I}_{D1} < 0$ at low $Q^2$.
Since, as mentioned, the large-$Q^2$ relation is 
equivalent to ${\cal I}_S = 2 b \, {\cal I}_{D1}$,
and ${\cal I}_S >0$ does not change sign, 
the inference is that $b \, {\cal I}_{D1}$ 
should change sign at large $Q^2$
in order to satisfy (\ref{eqGM3extra}).
We recall, however, as discussed, 
that $b \, {\cal I}_{D1}$ 
does not change sign in the present range of study.

The corollary of the previous discussion is that the parametrizations 
(\ref{eq-psiS})--(\ref{eq-psiD1}) are incompatible 
with (\ref{eqGM3extra}), when the parameters of the 
radial wave functions are determined by 
available lattice QCD data. 
We can enforce the verification of the condition (\ref{eqGM3extra})
but then we fail to obtain an accurate description
of the spacelike form factor data.

To ensure the validity of Eq.~(\ref{eqGM3extra}),
we consider then the parametrizations (\ref{eq-psiSb})--(\ref{eq-psiD1b}).
The new expression for $\psi_S$  is compatible with a change of sign 
on the overlap integral ${\cal I}_S$.
We modify also the expressions for $\psi_{D3}$ and $\psi_{D1}$
including a factor common to $\psi_S$.
The expression for $\psi_{D1}$ is motivated by 
the relation between the overlap integrals  ${\cal I}_S$ and ${\cal I}_{D1}$.
As for $\psi_{D3}$, we use a dependence analog to $\psi_{D1}$ 
for consistency with the $D$-state structure.
The parametrizations (\ref{eq-psiSb})--(\ref{eq-psiD1b}) 
are also compatible with the relations (\ref{eqInt-infty})~\cite{Note2}.
The new form for $\psi_S$
can be used to induce a change of sign in the function $G_{M1}$
and to ensure that  $G_{M1}$ and $G_{M3}$ have the same sign at large $Q^2$,
according to ${\cal I}_S =2  b\, {\cal I}_{D1}$.
The condition ${\cal I}_S = 2 b\, {\cal I}_{D1}$ 
may be difficult to impose analytically
but can be approximated numerically within 
a certain accuracy for a given region of $Q^2$,
centered on a given large scale $\bar Q^2$,
with a particular choice of parameters.

To the best of our knowledge, this is the first time that 
the the relation (\ref{eqGM3extra})
is considered in the context of the $\frac{3}{2}^+$ baryon form factors,
in general, and the $\Omega^-$ form factors in particular.

We tested tentatively if there was a scale $\bar Q^2$ 
where the large  $Q^2$ condition (\ref{eqGM3extra}) 
could be satisfied in the interval 
$[\bar Q^2 -\Delta Q^2,\bar Q^2 + 2 \Delta Q^2]$,
for a given  $\Delta Q^2$, 
with an accuracy better than $\epsilon$.
We consider the upper limit $Q_p^2= \bar Q^2 + 2 \Delta Q^2$,
with a difference $2 \Delta Q^2$ for $\bar Q^2$ to ensure 
a smoother convergence in the range between $ Q_m^2= \bar Q^2 -\Delta Q^2$
and $\bar Q^2 + \Delta Q^2$.

To include the  condition (\ref{eqGM3extra}) in our fit, 
we use the function  
\ba
{\cal R}(Q^2) =
\frac{G_{M1} (Q^2) - \frac{4}{5}\tau G_{M3} (Q^2)}{G_{M1}(Q^2)}.
\ea  
If the condition (\ref{eqGM3extra}) is valid 
we should have ${\cal R} (Q^2) \simeq 0$
for values of $Q^2$ near a large scale $\bar Q^2$.

\begin{table*}[t]
\begin{tabular}{l  c  c c c c c c c c}
\hline
\hline
    & $a$ & $b$ & $\alpha_1^\prime$ & $\alpha_2^\prime$ & $r_S$ & 
$\alpha_1$ & $\alpha_2$ & $\alpha_3$ & $\alpha_4$      \\
\hline
\hline
Fit SL/TL + LQ2         &   0.0316  & 0.2859 &  11.141  &  0.0818 & 1.845$\times$10$^{-3}$ & 
0.2018 & 0.1494 & 0.0789 & 1.811  \\  
\hline
\hline
\end{tabular}
\caption{\footnotesize 
Adjustable parameters of the fits to the spacelike and timelike (SL/TL) data
with the large-$Q^2$ (LQ2) constraint (\ref{eqGM3extra}),
based on the parametrization of the radial wave functions 
(\ref{eq-psiS})--(\ref{eq-psiD1}).
$S$ state ($r_S$, $\alpha_1^\prime$, $\alpha_2^\prime$, $\alpha_1$, $\alpha_2$),
$D3$ state ($\alpha_1$, $\alpha_2$, $\alpha_3$) 
and $D3$ state ($\alpha_1$, $\alpha_2$, $\alpha_4$).}
\label{tableParam2}
\end{table*}

We need then to test numerically if  
$|{\cal R} (Q^2)|$ is smaller than a given 
value in an interval $[ Q_m^2, Q_p^2]$.
Recall, however, that one expects also 
that ${\cal R} (Q^2) \propto 1/Q^2$, for large $Q^2$,
since $(G_{M1} - \frac{4}{5}\tau G_{M3} ) \propto 1/Q^6$
and $G_{M1} \propto 1/Q^4$, apart logarithmic corrections.
To take into account the falloff ${\cal R} (Q^2)  \propto 1/Q^2$,
we consider the condition 
\ba
|{\cal R} (Q^2)| < \frac{Q^2}{Q_p^2} \, \epsilon.
\ea
for $ Q_m^2 < Q^2 < Q_p^2$.
Thus, in the upper limit, one requires that 
$|{\cal R} (Q_p^2)| < \epsilon$.
For $Q^2 < Q_p^2$, however, one demands a softer 
condition in order to obtain 
a smooth convergence of  $G_{M1}$ to $ \frac{4}{5}\tau G_{M3}$ 
in the point $Q^2 = Q_p^2$, with the required precision.

When we implement the constraints described above,
the solutions (fits) compatible with 
the relation (\ref{eqGM3extra}) are characterized 
by the values of $\bar  Q^2$, $\Delta  Q^2$ and $\epsilon$.
To obtain a significant range 
of convergence, we choose $\Delta Q^2 = 100$ GeV$^2$.
Once $\bar  Q^2$ and  $\Delta  Q^2$ are defined,
the value of $Q_p^2$ corresponds to  
the upper limit of the parametrization.

We vary then the value of $\bar Q^2$ 
looking for solutions with an accuracy better than $\epsilon$.
The numeric calculations indicate that 
a fair description of the data  ($\chi^2 \le 2$)
with a 1\% accuracy ($\epsilon =0.01$) 
is obtained  for $\bar Q^2= 900$ and 1000 GeV$^2$.
Solutions with $\epsilon =0.02$ (2\% accuracy) 
can be obtained in a wider region of $\bar Q^2$, 
but the convergence between the two functions is not so smooth.
We consider then the solutions with $\epsilon =0.01$.

We choose  the solution with $\bar Q^2= 900$ GeV$^2$, 
since it provides the lowest  value for chi square
(best description of the overall data).
Solutions with large $\bar Q^2$ 
tend to provide a better description of the $|G(q^2)|$ data 
and to increase the values of $G_{M3}$ at low $Q^2$
(a less accurate description of the $G_{M3}$ data).
As one of the motivations of the present work 
is to investigate if there are solutions based 
on the our formalism compatible 
with the present $\Omega^-$ electromagnetic 
form factor data and with the condition (\ref{eqGM3extra}),
we do not try to fine tune the value of $\bar Q^2$.
In principle better solutions (lower chi square values)
can be obtained varying $\bar Q^2$ near 900 GeV$^2$.
Those solutions are, however, very similar to the selected one,
and qualitatively equivalent to the case $\bar Q^2= 900$ GeV$^2$.
Future experiments can provide further constraints
on the parametrizations of the  $\Omega^-$ form factors
and help to decide the appropriated scale for $\bar Q^2$.

In the context of our formalism the 
parametrizations  (\ref{eq-psiSb})--(\ref{eq-psiD1b})
are interpreted as effective corrections to 
the radial wave functions of the $\Omega^-$ baryon,
which resemble some properties of pQCD in a simplified form.

The parameters associated with the global fit 
of the spacelike and timelike data, 
constrained by the relation (\ref{eqGM3extra}),
are presented in Table~\ref{tableParam2}.
The values of chi square per data point are presented in 
the last row of Table~\ref{tableChi2} (Fit SL/TL + LQ2).

The best fit to the data, including 
the band of variation estimated from the values of $N^2$
is represented in Figs.~\ref{figFF1} and \ref{figFF2}
by the orange band.
The more significant differences to the previous fits 
can be observed in Fig.~\ref{figFF2} for $G_{E2}$ and $G_{M3}$.
We recall that, based on the previous discussion, 
the upper limit of the present estimates is $Q_p^2= 1100$ GeV$^2$.

The function $G_{E2}$ (orange band) is enhanced at low $Q^2$ compared
to the previous fits.
At $Q^2=0$, we obtain $G_{E2} (0)= 1.12 \pm 0.04$,
a larger value than the estimate 
obtained by the SL/TL fit, $G_{E2}(0) = 0.91 \pm 0.02$,
and almost twice the estimate from Ref.~\cite{Omega2}.
We recall that different extrapolations to $Q^2=0$
from the $G_{E2}$ data, are expected 
due to the large uncertainty of the data.
We avoid a detailed comparison of the results for $G_{E2}$
with the literature, since our results are determined by the 
global fit to the lattice QCD data, 
presented in the left panel of Fig.~\ref{figFF2}.

As for the function $G_{M3}$, we obtain a larger 
estimate for $G_{M3}(0)$ with $G_{M3}(0) =27.6 \pm  1.1$ (orange band).
The solution for  $G_{M3}$ is also characterized 
by a strong falloff with $Q^2$, which contributes 
to a reduction of the magnetic contribution to 
the effective form factor $|G(q^2)|$ at large $q^2$.

The result for the effective form factor $|G(q^2)|$, 
is represented in Fig.~\ref{figGT-Omega} by the solid line 
within the range represented by the orange band. 
This estimate is very close to the 
parametrization which does not take into account 
the large-$Q^2$ constraint (represented by the dash lines).
Overall, one can say that the estimates 
based on the covariant spectator quark model
are in good agreement with the large-$q^2$ data from CLOE
within the theoretical errors of the asymptotic estimate.

\begin{figure}[t]
\vspace{.5cm}
\centerline{
\mbox{\includegraphics[width=3.2in]{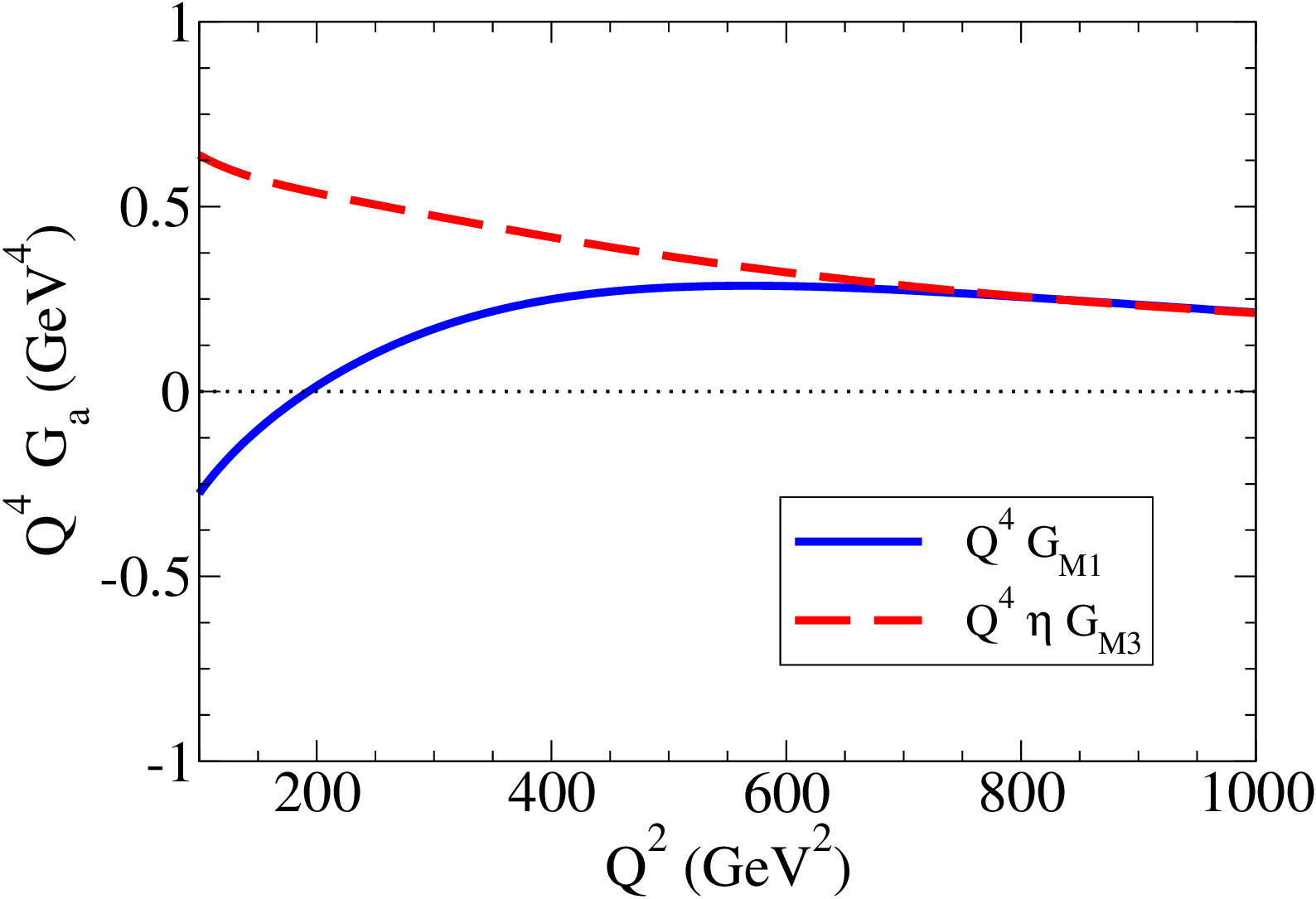} }}
\caption{\footnotesize{Test of 
the relation (\ref{eqGM3extra}) for large values of $Q^2$.
We use $\eta= \frac{4}{5} \tau$. 
}}
\label{fig-LQ2}
\end{figure}

Since the results of the last fit (SL/TL + LQ2) depend on 
the accuracy of the relation (\ref{eqGM3extra}),
we tested the convergence in the selected region (800--1000 GeV$^2$).
In Fig.~\ref{fig-LQ2}, we present the comparison 
of $G_{M1}$ and $\frac{4}{5} \tau G_{M3}$ multiplied by $Q^4$.
The factor $Q^4$ is included in order 
to remove the effects of the leading-order dependence 
of the form factors, 
dominated by terms of the order $1/Q^4$.
The results confirm that the relations 
are valid in good approximation (smooth convergence)
in the range 700-1000 GeV$^2$.

\begin{figure*}[t]
\vspace{.5cm}
\centerline{
\mbox{
\includegraphics[width=3.2in]{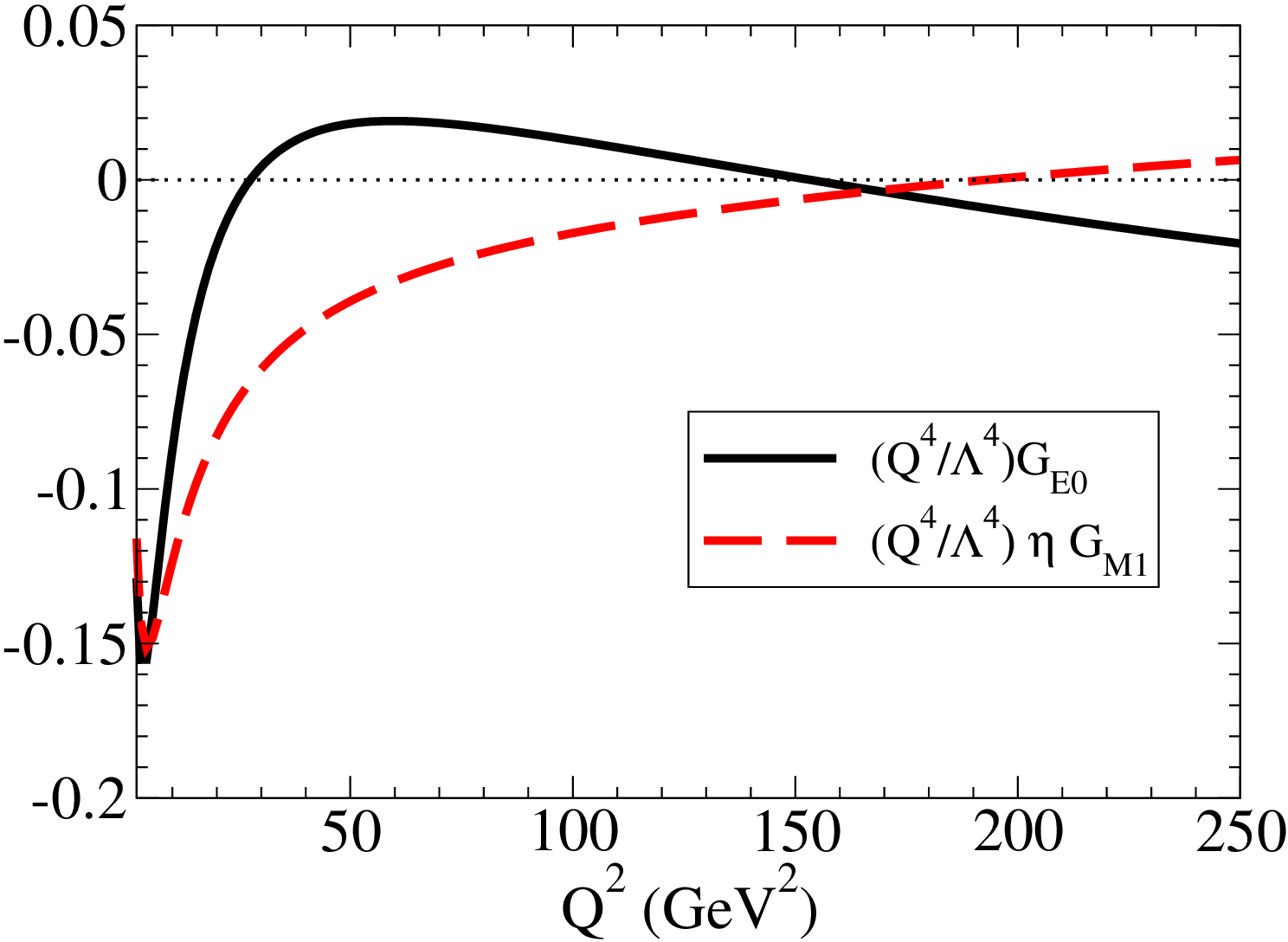} \hspace{.7cm}
\includegraphics[width=3.15in]{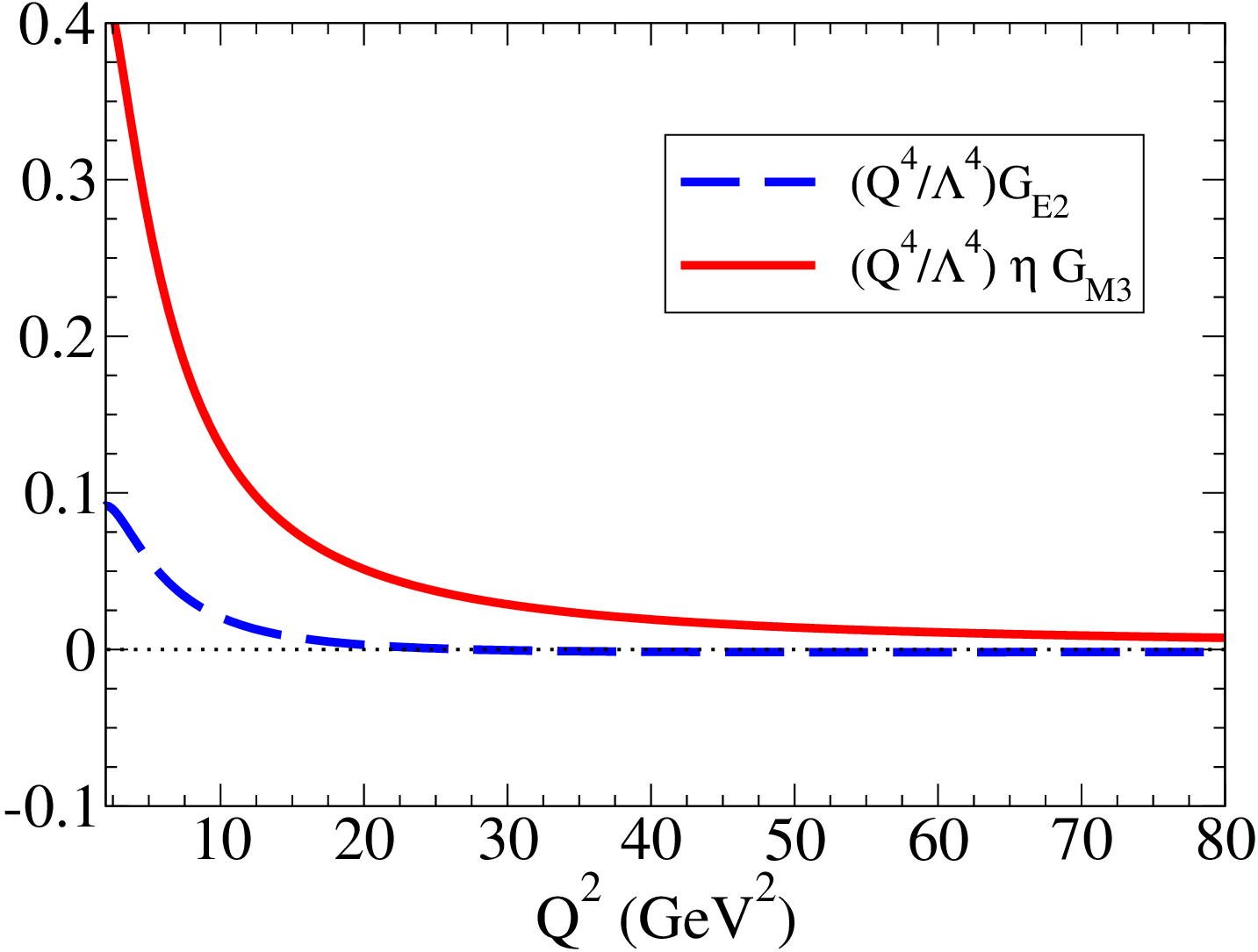}
}}
\caption{\footnotesize{Representation 
of the form factors $G_{E0}$,
$G_{M1}$, $G_{E2}$ and $G_{M3}$ for large $Q^2$,
calibrated by the factor  $\left(\frac{Q^2}{\Lambda^2}\right)^2$, 
with $\Lambda^2= 2$ GeV$^2$ (spacelike region).
We use also $\eta= \frac{1}{4}$ to reduce 
the scale of $G_{M1}$ and $G_{M3}$.
Notice the zeros of the functions $G_{E0}$, $G_{M1}$ and $G_{E2}$.
}}
\label{figFF3}
\end{figure*}

The deviation from a horizontal line indicates that, 
even in the range $Q^2 =700$--1000 GeV$^2$,
logarithmic corrections or terms of the order $1/Q^6$ are still meaningful.
We checked numerically that the 
form of the present parametrizations for 
the radial wave functions is consistent
with a very slow falloff of the correction 
to the leading order $1/Q^4$.
We do not attempt to check the exact scaling 
at very large $Q^2$, because the scale $Q^2 = 1100$ GeV$^2$ 
is interpreted as the upper limit of our calculations,
and also because we are well above the range 
of the present day experiments, from which 
our model parametrizations are calibrated.

To have an idea about the general behavior 
of the form factors above the region 
displayed on Figs.~\ref{figFF1} and \ref{figFF2},
we represent in Fig.~\ref{figFF3} the four form factors
for $Q^2  \ge 2$ GeV$^2$.
We multiply the functions 
by $\left(\frac{Q^2}{\Lambda^2}\right)^2$ 
with $\Lambda^2= 2$ GeV$^2$ 
in order to suppress (part of) the falloff of 
the form factors for large $Q^2$.
With this representation, we complement 
the results from Figs.~\ref{figFF1} and \ref{figFF2},
since $\left(\frac{Q^2}{\Lambda^2}\right)^2 = 1$, 
at the threshold of the representation. 
The inflection points near the threshold in the graph for $G_{E0}$
and $G_{M1}$ are the consequence 
of the factor $Q^4$ and are not relevant for the present discussion.
The zero for the functions $G_{E0}$ 
and $G_{E2}$ for $Q^2 = 28$ GeV$^2$ is a consequence of the 
of the zero of the function $\tilde g_\Omega$
defined by Eq.~(\ref{eq-gtilde}) and 
the particular parametrization 
of the strange quark form factors $f_{i0}$,
from Eqs.~(\ref{eqQff1}) and (\ref{eqQff2}), 
determined in Ref.~\cite{Omega} by the study 
of the decuplet baryon form factors.
This result is independent of the 
overlap integrals ${\cal I}_{S}$ and ${\cal I}_{D3}$.
Above $Q^2 = 28$ GeV$^2$, $G_{E2}$ became negative 
but almost negligible. 
The change of signs on  $G_{E0}$  for $Q^2 \simeq 153$ GeV$^2$
and of $G_{M1}$ for $Q^2 \simeq 193$ GeV$^2$ 
is the consequence of the parametrizations
(\ref{eq-psiSb})--(\ref{eq-psiD1b}),
more specifically of the decomposition of $\psi_S$ into two terms.
The changes of sign on $G_{M1}$ are necessary for the validity 
of the relation (\ref{eqGM3extra}),
as discussed already.

In the case the zero on functions $G_{E0}$ 
and $G_{E2}$ is confirmed by future experiments,
the zero is justified by a mechanism similar 
to the one associated with the zero of the 
proton electric form factor 
$G_{Ep} = F_{1p} - \frac{Q^2}{4 M_N^2} F_{2p}$~\cite{Nucleon,Gayou02,Puckett17},
where there is a competition between the two terms
(Dirac $F_{1p}$ and Pauli $F_{2p}$ form factors).

Only new data (empirical or lattice) can 
confirm if our parametrization of the 
strange quark form factors, 
calibrated at low $Q^2$~\cite{Omega}, 
can be extended very large values of $Q^2$.



\subsection{Discussion of the results}

From the results for the parametrizations 
which describe the SL/TL data
and the parametrizations which verify also the LQ2 condition, 
one can conclude that it is possible to derive parametrizations 
which take into account the present knowledge 
of the $\Omega^-$ form factors in the spacelike region 
(mainly lattice QCD data for $G_{E0}$, $G_{M1}$ and $G_{E2}$),
as well as the timelike data (effective form factor data).

We notice, however, that the description 
of the function $G_{M3}$ is not as accurate 
as the others ($G_{E0}$, $G_{M1}$, and $G_{E2}$)
and that the description of the effective form factor $|G(q^2)|$
is also not as perfect as the first 
three multipole form factors.

The limitations in the description of 
the functions  $G_{M3}$ and  $|G(q^2)|$ are, however, 
also related to the limited weight of those functions 
in the fit.
Notice that one has 100 spacelike data points for 
$G_{E0}$  $G_{M1}$ and $G_{E2}$,  
only one point for $G_{M3}$, and two points for $|G (q^2)|$.
The fit is then dominated by the 100 points 
of the first three form factors ($G_{E0}$  $G_{M1}$ and $G_{E2}$).
In principle, we could enhance the impact of the points 
$G_{M3}$ and  $|G(q^2)|$ including an extra weight in the fit
(equivalent to a reduction on the error bars), 
but the procedure goes against the principle of 
giving the same treatment to the spacelike and the timelike regions
(equal weight to spacelike and timelike points).

From the comparison between the two global fits
(with and without LQ2) one can conclude that the consideration 
of the large-$Q^2$ relation (\ref{eqGM3extra}) 
decreases the accuracy of the final fit, 
at the expenses of a less accurate description of $G_{M3}$,
as can be observed in Table~\ref{tableChi2} 
(compare values of the column $G_{M3}$). 
Our estimate of $G_{M3}$ deviates from the lattice data 
by 2.3 standard deviations (see Fig.~\ref{figFF2}). 

One can notice, however, that the estimate of $G_{M3}$ 
depends exclusively on a unique point estimated 
in 2009, affected with a large uncertainty.
The lattice QCD techniques used in the past 
can be used today to estimate $G_{M3}$ 
for a larger set of $Q^2$ points with a better accuracy.
New lattice QCD simulations of $G_{M3}$ 
for values of $Q^2$ closer to $Q^2=0$ can then 
be used to test our estimate for $G_{M3}(0)$ 
and the shape of  $G_{M3}$  at low $Q^2$.

Our best parametrization (SL/TL + LQ2) suggests then 
that our estimate for $G_{M3}(0)$
is most likely incompatible with the available lattice result 
for $G_{M3}(0.23 \; \mbox{GeV}^2)$.
The large values of the function $G_{M3}$ are a consequence 
of the consideration of the timelike data and 
the large-$Q^2$ condition (\ref{eqGM3extra}).
It is interesting to notice, however,
that our large estimate for $G_{M3}(0)$ is comparable with 
other theoretical estimates presented 
in the literature as discussed next.

In the literature, there are only a few estimates 
of the function $G_{M3}$.
In chiral perturbation theory, 
the magnetic octupole momentum vanishes
at the next-to-leading order 
of the chiral expansion~\cite{Arndt03}.
Also the chiral solition model approaches give null 
results for the function $G_{M3}$~\cite{Kim19}.
Calculations based on the 
Dyson-Schwinger equations formalism~\cite{Nicmorus10,Sanchis13} 
suggest that the magnitude of $G_{M3}$ may 
be small but the sign is uncertain.
Reference~\cite{Nicmorus10} estimates 
that $G_{M3}(0) \simeq + 0.5$, 
while the calculations from Ref.~\cite{Sanchis13}
point to $G_{M3} (0) \simeq - 0.3$ or   $G_{M3} (0) \simeq - 0.5$.
The Dyson-Schwinger results of the function $G_{M3}$ 
are sensitive to the truncation used 
in the interaction kernels~\cite{Sanchis13}.

In Table~\ref{tableGM3}, we compare our estimates
with calculations based on QCD sum rules~\cite{Aliev09}
and with the $SU(3)$ non covariant quark model~\cite{Buchmann08}.
The first estimate from Ref.~\cite{Buchmann08} 
is based on an exact $SU(3)$ model (symmetric wave function);
the second estimate considers the breaking of $SU(3)$. 
From the observation of the table, 
we can conclude that our estimate is 
comparable in magnitude with the estimates 
based on QCD sum rules and the symmetrical 
non covariant quark model.
Table~\ref{tableGM3} includes also the results for the 
octupole magnetic moment for an easy comparison 
with other works.

The previous analysis shows that the estimate 
of the octupole magnetic moment is an open problem, 
and more theoretical and experimental efforts 
are necessary to clarify the situation.
Also for that reason, 
lattice QCD simulations 
for $G_{M3}$ at the physical mass of the $\Omega^-$ are mandatory.
Accurate results for $G_{M3}$ can be used to extrapolate 
$G_{M3}(0)$, following the lines of previous estimates 
of $G_{E2}(0)$ based on lattice QCD data~\cite{Alexandrou10,Omega2}.

One can also discuss the shape of 
$\Omega^-$ based on the electric quadrupole 
[proportional $G_{E2} (0)$]
and magnetic octupole 
[proportional $G_{M3} (0)$] moments,
which measure deviations from the electric charge 
and magnetic dipole distributions from a spherical form.
For negative charge baryons, positive moments 
indicate a compression along the polar axis 
(oblate shape), and negative moments 
indicate an elongation in the polar axis (prolate shape),
since the charge density depends 
on the baryon charge~\cite{Deformation,Alexandrou09,Buchmann01}.
We conclude, then, that the $\Omega^-$ 
presents a distribution of electric charge 
and magnetic dipole compressed along the spin axis,
corresponding to an oblate shape.
The shape of the $\Omega^-$ resembles then the shape of the $\Delta^+$ 
[positive charge, $G_{E2} (0) <0$ and $G_{M3} (0) <0$],
according to the covariant 
spectator quark model estimates~\cite{DeltaFF},
ignoring the scale of the deformation.

Based on our results for $G_{E0}$, we estimate 
the size of the $\Omega^-$ given by  
the electric charge square radius as
$r_{E0}^2 \simeq 0.24$ fm$^2$.
This result corresponds to about one-third 
of the value measured for the proton
($r_{Ep}^2 = 0.707$ fm$^2$)~\cite{PDG20},
suggesting that $\Omega^-$ is much more compact than the proton.

\begin{table}[t]
\begin{tabular}{l  c  c }
\hline
\hline
    & $G_{M3}(0)$ & ${\cal O}_\Omega$ ($10^{-3}$ fm$^3$)       \\
\hline
\hline
QCD Sum Rules~\cite{Aliev09} &  64.3$\pm$16.1 &  16.0$\pm$4.0\\
Non Covariant QM (sym)~\cite{Buchmann08}\spQ \spQ & 48.2 &  12.0\\
Non Covariant QM~\cite{Buchmann08}     & 12.2 &  3.04 \\[.5cm]
Spectator~\cite{Omega2} &  15.5  &  3.85 \\
Spectator SL          &  14.4$\pm$0.5  &  3.61$\pm$0.14 \\
Spectator SL/TL       &  21.0$\pm$0.6  &  5.22$\pm$0.14 \\
Spectator SL/TL + LQ2 &  27.6$\pm$1.1  &  6.88$\pm$0.27 \\
\hline
\hline
\end{tabular}
\caption{\footnotesize 
Results for $G_{M3}(0)$ and magnetic octupole in units $10^{-3}$ fm$^3$.
An alternative representation of the magnetic octupole 
uses units $e\,$fm$^3$, where is the elementary electric charge 
($e= \sqrt{4\pi \alpha} \simeq 0.303$).
Non Covariant QM (sym) indicates a $SU(3)$ symmetrical model.}
\label{tableGM3}
\end{table}

Concerning the results for the function  $|G (q^2)|$, 
in the timelike region,
we conclude that the estimates 
which exclude and include the condition (\ref{eqGM3extra})
are very similar (see Fig.~\ref{figGT-Omega}).
The main differences between the two parametrizations
appear for the function $G_{M3}$, 
providing an additional motivation 
for performing lattice QCD simulations for this form factor.

Two final remarks about about our calculations 
in the timelike region and the range of validity 
of the model calculations are in order.

The $|G (q^2)|$ data from CLEO used in our calibrations 
correspond to the values of $q^2=14.2$ and 17.4 GeV$^2$.
These values for $q^2$ are  still very close to the threshold of 
the $e^+ e^- \to \Omega^- \bar \Omega^+$ reaction 
($q^2= 4 M_\Omega^2 \simeq 11.2$ GeV$^2$).
Therefore, the use of an expression valid for very large $|Q^2|$
may not be fully justified.
Since theses are the only available data for $q^2 > 0$,
we used the data anyway.

The second remark is that, for simplicity, 
we did not take into account the uncertainty 
associated with our estimate of $|G (q^2)|$, 
in the calibration of the model in 
the calculation of the chi square.
If we take into account those uncertainties  
(integrating out on a normal distribution centered 
on the model result for  $|G (q^2)|$),
we improve the quality of the description of this function,
since the data are inside the one-standard-deviation region.
In that case, we would obtain a smaller value 
for the $|G|$ chi square, and consequently a smaller 
global chi square per data point.
We avoid this procedure because, as mentioned above, 
the theoretical estimates 
are still far way from the large-$|Q^2|$ region, 
where the estimate and the estimate of error are valid,
and also because we do not want to 
reduce at this stage the impact of 
the timelike data in the calibration of the model.

Future measurements for larger $q^2$ values,  
farther away from the threshold $4 M_\Omega^2$, 
are of capital importance to check if the present  
falloffs associated to $|G (q^2)|$ and $G_{M3}$ are correct 
or if the present shape of the function $G_{M3}$ 
has to be corrected.



\subsection{Predictions of $|G|$ for large $q^2$}

Our final parametrization, 
which takes into account the timelike region and 
the large-$Q^2$ relation (\ref{eqGM3extra}),
can now be used to make predictions for  
effective form factor $|G (q^2)|$, for larger values of $q^2$.
Recall that the function  $|G (q^2)|$ is extracted from
$e^+ e^-$ collision experiments cross sections, 
which can be measured presently in several laboratories, 
for several hyperons, including 
the $\Omega^-$~\cite{BaBar,Dobbs17a,BESIII,Dobbs14a,Singh17a,Hyperons}.

Our estimates are presented in Table~\ref{tableG}.   
These predictions can be compared with future measurements 
of the effective form factor $|G (q^2)|$ of the $\Omega^-$ baryon.
\begin{table}[h] 
\begin{tabular}{l  c }
\hline
\hline
$q^2$ \spQ&  \spQ  \spQ$|G (q^2)|$  ($10^{-3}$) \spQ \spQ  \\
\hline
\hline
 20  &   4.02  (3.51)\\
 25  &   2.75  (1.91)\\
 30  &   1.97  (1.14)\\
 35  &   1.47  (0.73)\\ 
 40  &   1.14  (0.50)\\
 45  &  0.900  (0.350)\\
 50  &  0.728  (0.255)\\
 55  &  0.600  (0.191)\\
 60  &  0.502  (0.147)\\
 65  &  0.426  (0.115)\\
 70  &  0.365  (0.091)\\
 75  &  0.316  (0.074)\\
 80  &  0.276  (0.060)\\
\hline
\hline
\end{tabular}
\caption{\footnotesize Estimates of  $|G (q^2)|$ for large $q^2$, 
compatible with condition (\ref{eqGM3extra}).
The values between comas represent the 
average of the upper and lower limits.
Those estimates are justified above $q^2= 40$ GeV$^2$ 
(symmetric deviations).
Below 40 GeV$^2$ the upper limit dominates over 
the lower limit (unsymmetrical error bars).}
\label{tableG}
\end{table}

\section{Conclusions}
\label{secConclusions}

In the last few years, there have been important developments 
in the experimental study of the electromagnetic structure 
of baryons in the timelike region, 
based on $e^+ e^- \to B \bar B$ reactions,
for several baryon ($B$) states. 
Among other observables, one has access to 
the effective electromagnetic form factor $|G(q^2)|$
of the $\Omega^-$ baryon in the region ($q^2 \ge  4 M_\Omega^2$).

The recent measurements of the $\Omega^-$ timelike
effective form factor are very pertinent  
because there is no experimental information 
about the  $\Omega^-$ electromagnetic structure,
apart from the charge and the magnetic moment.
There are,  however, lattice QCD simulations, 
at the strange quark physical mass, 
which can be interpreted as a reliable representation 
of the physical $\Omega^-$ (reduced light quark effects).
Those lattice QCD simulations are performed 
in the spacelike region ($Q^2= -q^2 \ge  0$) 
for the electric charge ($G_{E0}$),
magnetic dipole ($G_{M1}$) and electric quadrupole ($G_{E2}$)
form factors.
For the magnetic octupole ($G_{M3}$) form factor 
there are only simulations with large errors.

In the present work, we propose using 
the available information about the $\Omega^-$
to obtain a complete picture 
of the electromagnetic structure of the $\Omega^-$,
in the spacelike region ($q^2 \le 0$),
as well as in the timelike region ($q^2 \ge 4 M_\Omega^2$).
We combine the experimental information 
(magnetic moment, $q^2=0$), with 
the lattice QCD data in the spacelike region, 
and the effective form factor $|G(q^2)|$ in the timelike region.

We also take into account constraints from pQCD,
requiring that $G_{E0}$, $G_{M1} \propto 1/Q^4$ and 
$G_{E2}$, $G_{M3} \propto 1/Q^6$ for very large $Q^2$
(apart logarithmic corrections). 
In addition, we consider also a condition 
which relates the $G_{M1}$ and $G_{M3}$ form factors 
in the asymptotic region.
The relation $G_{M1}= \frac{4}{5} \tau G_{M3}$  
has a strong impact on the dependence 
of the form factors on $Q^2$ for very large $Q^2$.
The discussion of the relevance of the correlation 
between the from factors $G_{M1}$ and $G_{M3}$ 
is an important contribution of the present work.
As far was we know, this relation has not been 
discussed in the literature.
In the present work, the correlation between 
the magnetic form factors is manifest for 
a scale of $Q^2$ around 900 GeV$^2$.

Our analysis is based on the covariant spectator quark model formalism,
where the $\Omega^-$ is a combination of an $S$ state and two $D$ states.
The $D$ states are responsible by the
deformation of the $\Omega^-$ system
(nonzero results for $G_{E2}$ and $G_{M3}$).
Part of the model (strange quark form factors) 
was calibrated by previous studies of the decuplet baryon.
The parameters related to the $S$- and $D$-state 
radial wave functions and the admixture parameters 
are calibrated by the available physical and lattice QCD data
for the $\Omega^-$. 
For the calculations in the timelike region, 
we use asymptotic relations valid for very large $|Q^2|$.
The existing data are 
described by a small (0.1\%) $D$-state 
contribution associated with total quark spin 3/2,
and a more significant (7.6\%) $D$-state 
contribution associated with total quark spin 1/2.
The first contribution dominates $G_{E2}$,
the second contribution dominates $G_{M3}$.
Our results suggest a distribution of
charge and magnetism compressed along the spin axis,
corresponding to an oblate shape of the $\Omega^-$ baryon.

We conclude that the available information on the $\Omega^-$ baryon,
including the lattice QCD data in the spacelike region
and the $|G(q^2)|$ physical data in the timelike region, 
is compatible with a large value for $G_{M3}$ 
near $Q^2=0$ [$G_{M3} (0) \simeq 28$] 
and a fast falloff of $G_{M3}$ for \mbox{large $|Q^2|$.}
It is worth noticing, however, that the present estimations 
are based on lattice QCD simulations, 
limited in the range of $Q^2$ and precision in the spacelike region,
on 2 timelike data points, and a single lattice calculation 
of $G_{M3}$  with a large uncertainty.
More timelike data at larger $q^2$ are necessary 
to better constrain the shape of the form factors 
at large $|Q^2|$ and to increase the impact 
of the timelike data in the model calibrations.

The value of $G_{M3} (0)$ is, presently, an open question.
The available lattice QCD data 
are compatible with positive and negative values.
Our estimate suggests that $G_{M3} (0)$ is large 
and positive, consistent with estimates based on 
$SU(3)$ quark models and QCD sum rules.
Some frameworks, however, point to a small $G_{M3}$.
Accurate simulations possible with the 
present state-of-the-art lattice QCD techniques 
are necessary to better infer the magnitude of $G_{M3}(0)$,
and the shape of the function $G_{M3}$ 
at low and intermediate $Q^2$.

Future accurate lattice QCD simulations 
of the four $\Omega^-$ form factors at 
low and intermediate $Q^2$ 
and $e^+ e^- \to \Omega^- \bar \Omega^+$ collision experiments 
for large values of $q^2$ (say $q^2 > 30$ GeV$^2$)
are very important to determine the shape 
of the $\Omega^-$ form factors and test 
the present predictions.
Those results can also confirm if the model calibrations derived 
from low-$Q^2$ lattice QCD data 
are also valid in the large-$Q^2$ region.

\begin{acknowledgments}
G.R.~was supported by the Funda\c{c}\~ao de Amparo \`a
Pesquisa do Estado de S\~ao Paulo (FAPESP) --
Project No.~2017/02684-5, Grant No.~2017/17020-BCO-JP.
\end{acknowledgments}

\appendix

\setcounter{table}{0}
\renewcommand{\thetable}{A\arabic{table}}

\section{Perturbative QCD estimates for the $\Omega^-$ form factors}
\label{app-PQCD}

We discuss here the asymptotic expressions 
associated with the form factors $G_{E0}$, $G_{M1}$, $G_{E2}$, and $G_{M3}$.
The derivation of those expressions follows 
the analysis of C.~Carlson \etal~\cite{Carlson0} 
of the helicity transition amplitudes for very large $Q^2$,
extended here to $\frac{3}{2}^+$ baryons.

\subsection{Electromagnetic current 
and form factors}

The current $J^\mu$ associated with the 
$\gamma^\ast \Omega^- \bar \Omega^+$ vertex
can be written in the  
Lorentz-invariant and gauge-invariance 
form~\cite{Alexandrou09,DeltaFF,Nozawa90} 
\ba
J^\mu & = &
- \left[ F_1^\ast (Q^2) g^{\alpha \beta} + F_3^\ast (Q^2) \frac{q^\alpha q^\beta}{4M^2}
\right] \gamma^\mu \nonumber \\
& & - 
\left[ F_2^\ast (Q^2) g^{\alpha \beta} + F_4^\ast (Q^2) \frac{q^\alpha q^\beta}{4M^2}
\right] \frac{i \sigma^{\mu \nu}}{2 M},
\label{eqJmu1}
\ea
in elementary charge units ($e$).
The value of the charge ($e_{\Omega}=-1$) is included 
on the function $F_1^\ast$  
[$F_1^\ast(0) = G_{E0} (0) = -1$].

In the previous expressions $q$ is the photon momentum,
$M$ is the baryon mass, and $F_i^\ast$ ($i=1,...,4$) 
are structure form factors.
In the following, we use {\it elementary} 
form factors to identify $F_i^\ast$.
In Eq.~(\ref{eqJmu1}), the free indices ($\alpha$, $\beta$) 
are contracted with Rarita-Schwinger spinors 
$\bar u_\alpha(P_+,S_z^\prime)$ and  $u_\beta(P_-,S_z)$
associated with the final and initial states
(spin projections $S_z^\prime$ and $S_z$), respectively.

The multipole form factors, $G_{E0}$, $G_{M1}$, $G_{E2}$, and $G_{M3}$
can be expressed as linear combinations of the elementary form factors.
For the following discussion, however, it is more convenient 
to define the auxiliary form factors $G_{i}$ ($i=1,..,4$)
\ba
G_1  &= F_1^\ast  - \tau  F_2^\ast , 
\label{eqG1} \\
G_2  &= F_1^\ast  + F_2^\ast , \\
G_3  &= F_3^\ast  - \tau  F_4^\ast , \\
G_4  &= F_3^\ast  + F_4^\ast.
\label{eqG4}
\ea

Using the new notation, we can write~\cite{Alexandrou09}
\ba
G_{E0} & =&  \left( 1 + \sfrac{2}{3} \tau \right)  G_1
- \frac{1}{3}  \tau( 1 + \tau) G_3, 
\label{eqGE0-0}
\\
G_{M1}  &=& \left( 1 + \sfrac{4}{5} \tau \right)  G_2
- \frac{2}{5}  \tau( 1 + \tau) G_4, \\
G_{E2} & =&  G_1
- \frac{1}{2} ( 1 + \tau) G_3, \\
G_{M3} &=& 
G_2
- \frac{1}{2} ( 1 + \tau) G_4.
\label{eqGM3-0}
\ea

The inverse relations are
\ba
F_1^\ast &=& \frac{1}{1 + \tau} ( G_1 + \tau G_2), 
\label{eqF1} \\
F_2^\ast &=& - \frac{1}{1 + \tau} ( G_1 - G_2), \\
F_3^\ast &=& \frac{2}{1 + \tau} (G_3  + \tau G_4 ), \\
F_4^\ast &=& - \frac{2}{1 + \tau} (G_3  - G_4),
\label{eqF4}
\ea
and
\ba
G_1 &= &  G_{E0} - \frac{2}{3} \tau G_{E2}, \label{eqG1a} \\
G_2 &= &  G_{M1} - \frac{4}{5} \tau G_{M3}, \label{eqG2a} \\
G_3 &= &  \frac{2}{1 + \tau} G_{E0} -
\frac{2}{1 + \tau} \left(1 + \sfrac{2}{3}\tau \right) G_{E2}, \label{eqG3a} \\
G_4 &= & 
 \frac{2}{1 + \tau} G_{M1} -
\frac{2}{1 + \tau} \left(1 + \sfrac{4}{5}\tau \right) G_{M3} .
\label{eqG4a} 
\ea


\subsection{Breit frame transition amplitudes}

The study of the asymptotic behavior 
of the electromagnetic form factors is simplified 
when we calculate the transition amplitudes 
between the possible spin projections in a given frame.
Following C.~Carlson, we consider the 
amplitudes in the Breit frame~\cite{Carlson0}
\ba
G_+  &= &\left< J^+ \right>_{+ \sfrac{1}{2}, - \sfrac{1}{2}}, 
\label{eqGp}
\\
G_{01} &=& \left< J^0 \right>_{+ \sfrac{1}{2}, + \sfrac{1}{2}}, 
\label{eqG01} \\
G_{03} &=& \left< J^0 \right>_{+ \sfrac{3}{2}, + \sfrac{3}{2}},
\label{eqG03}  \\
G_{-} &=& \left< J^+ \right>_{+ \sfrac{3}{2}, + \sfrac{1}{2}}, 
\label{eqGm}
\ea
where 
\ba
& &\left< J^0 \right>_{S_z^\prime, S_z}  
= \left< +  \sfrac{1}{2}{\bf q},  S_z^\prime 
\right| J^0 \left| -\sfrac{1}{2}{\bf q} ,S_z\right>, \\
& &\left< J^+ \right>_{S_z^\prime, S_z}  
= \left< +  \sfrac{1}{2}{\bf q},  S_z^\prime 
\right| J^+ \left| -\sfrac{1}{2}{\bf q} ,S_z\right>,
\ea
and $J^+ = - \frac{1}{\sqrt{2}}(J^1 + i J^2)$.
Note that in the Breit frame the initial 
state has helicity $-2 S_z$ and the final-state 
helicity $2 S_z^\prime$.

The notation $G_ \ell$ ($\ell =+,01,03,-$) 
is inspired on the notation from Refs.~\cite{Carlson0}
for the transitions between spin-1/2 states 
to spin-1/2 or spin-3/2 states 
(like the $\gamma^\ast N \to \Delta$ transition),
where $\ell =\pm,0$ are related to 
the photon polarization vectors.
In the present case, however, it is necessary 
to distinguish between two scalar transitions: 
$S_z= +\frac{1}{2} \to S_z^\prime = +\frac{1}{2}$, 
labeled as $G_{01}$, and 
$S_z= +\frac{3}{2} \to S_z^\prime = +\frac{3}{2}$, 
labeled as $G_{03}$. 
The amplitude $G_{03}$ is exclusive of the 
spin-3/2 elastic transitions.

\begin{table*}[t]
\begin{tabular}{l | c c c c | c c c c}
\hline
\hline
    &\spQ  $G_1$ \spQ & \spQ $G_2$ \spQ &\spQ  $G_3$ \spQ &\spQ $G_4$  \spQ 
    & \spQ $F_1^\ast$ \spQ  &  \spQ $F_2^\ast$ \spQ  & \spQ $F_3^\ast$ \spQ   & \spQ $F_4^\ast$ \spQ  \\
\hline
Case 1
     & $\frac{1}{Q^4}$  &  $\frac{1}{Q^6}$ &  
      $\frac{1}{Q^6}$  &  $\frac{1}{Q^8}$ &  
      $\frac{1}{Q^6}$  &  $\frac{1}{Q^6}$ &  
      $\frac{1}{Q^8}$  &  $\frac{1}{Q^{8}}$ \\
Case 2    & $\frac{1}{Q^4}$  &  $\frac{1}{Q^4}$ &  
      $\frac{1}{Q^6}$  &  $\frac{1}{Q^6}$ &  
      $\frac{1}{Q^4}$  &  $\frac{1}{Q^6}$ &  
      $\frac{1}{Q^6}$  &  $\frac{1}{Q^8}$   \\
\hline
\hline
\end{tabular}
\caption{\footnotesize 
Expected falloffs according with different large-$Q^2$ conditions.
{\bf Case 1:} consequence of the conditions
(\ref{eqGE0-as}),  (\ref{eqGE2-as}), and  (\ref{eqGM1-infty}).
{\bf 
Case 2:} ignore the condition (\ref{eqGM1-infty}).}
\label{tableFF-largeQ2}
\end{table*}

The Eqs.~(\ref{eqGp})--(\ref{eqGm}) represent 
the four independent, non-zero amplitudes.
The omitted cases can be related to these four cases.

The explicit calculation of the amplitudes  (\ref{eqGp})--(\ref{eqGm})
gives~\cite{Alexandrou09}
\ba
G_+ &= & 
-  \frac{2\sqrt{2}}{3} \sqrt{\tau} 
\left[ G_{M1} + \frac{6}{5} \tau G_{M3} \right], 
\label{eqGp-1}
\\
G_{01} &= &  G_1 + \frac{4}{3} \tau G_{E2}  =
G_{E0} + \frac{2}{3} \tau G_{E2}, 
\label{eqG01-1}\\ 
G_{03} &= &  G_1 = G_{E0} - \frac{2}{3} \tau G_{E2},
\label{eqG03-1}  \\
G_- &= & 
-  \sqrt{\frac{2}{3}}  \sqrt{\tau} G_2. 
\label{eqGm-1}
\ea
In the calculations, we use the normalization  
$\bar u_\alpha (P, S_z^\prime) u^\alpha 
(P,S_z) = - \delta_{S_z^\prime S_z}$
in the limit $P=(M,{\bf 0})$.

\subsection{Large-$Q^2$ behavior}

One can now use the pQCD analysis
from C.~Carlson \etal~\cite{Carlson0,Carlson}.
In the leading-order amplitude ($G_+$), 
the electromagnetic interaction 
preserves the helicity of the initial state 
(helicity conservation),
meaning that there is no spin flip of any quark. 
In the remaining amplitudes, the 
helicity is modified by the spin flip of one or more quarks.
Each spin flip transition is suppressed by a factor $m_q/Q$, 
where $m_q$ is the current quark mass~\cite{Carlson0}.
The consequence of this suppression is that 
$G_+$ has the slowest falloff with $Q^2$ for large $Q^2$,
and the remaining amplitudes are 
further suppressed by multiple factors  $m_q/Q$.

The asymptotic behavior of $G_{\ell}$ 
can be ordered as
\ba
G_+ & \propto & \frac{1}{Q^3}, 
\label{eqGp-as}\\
G_{01} & \propto & \frac{1}{Q^4}, 
\label{eqG01-as}
\\
G_{03} & \propto & \frac{1}{Q^4},
\label{eqG03-as} \\
G_- & \propto & \frac{1}{Q^5}. 
\label{eqGm-as}
\ea  
The amplitudes $G_{01}$ and $G_{03}$ have the same behavior because 
they are associated with the same helicity variation 
between the initial and the final state.

The pQCD constraints on the transition form factors 
can now be derived based on the comparison  
of the general relations (\ref{eqGp-1})--(\ref{eqGm-1})
and the asymptotic expressions  (\ref{eqGp-as})--(\ref{eqGm-as}).
We start our analysis with the relation for $G_-$.
From the comparison between  
(\ref{eqGm-1}) with the 
asymptotic result  (\ref{eqGm-as}), we conclude that
\ba
& &
G_2 = G_{M1} - \frac{4}{5} \tau G_{M3}  = {\cal O} 
\left( \frac{1}{Q^6} \right), 
\label{eq-G2-extra}
\ea

From the relations for $G_{01}$ and $G_{03}$, 
one can conclude from the comparison 
of the Eqs.~(\ref{eqG01-1}) and (\ref{eqG03-1})
with the relations  (\ref{eqG01-as}) and  (\ref{eqG03-as}),
respectively, that
\ba
G_{E0} \pm \frac{2}{3} \tau G_{E2} = 
 {\cal O} 
\left( \frac{1}{Q^4} \right), 
\ea 
meaning that
\ba
G_{E0}, \; \tau G_{E2} \propto \frac{1}{Q^4}.
\label{eq-extra1}
\ea

To finish the analysis, we look for the amplitude $G_+$.
Using Eqs.~(\ref{eqGp-1}),  (\ref{eqGp-as}) and 
(\ref{eq-G2-extra}), one obtains
\ba
G_+ &= & 
- \frac{5 \sqrt{2}}{3} \sqrt{\tau}
 \left[G_{M1}  +  {\cal O} 
\left( \frac{1}{Q^6}\right)  \right] \propto \frac{1}{Q^3}. 
\nonumber\\
\ea
from which we conclude that
\ba
G_{M1} \propto \frac{1}{Q^4}.
\label{eq-extra2}
\ea
Combining this result with (\ref{eq-G2-extra}),
we conclude also that
\ba
\tau G_{M3} \propto \frac{1}{Q^4}.
\label{eq-extra3}
\ea

A corollary of the relations (\ref{eq-extra1}), 
(\ref{eq-extra2}) and  (\ref{eq-extra3})
is that $G_{E0} \propto \tau G_{E2} \propto 1/Q^4$ 
and $G_{M1} \propto \tau G_{M3} \propto 1/Q^4$.

\subsection{Summary of the pQCD limit}

We can now summarize the expected results for the 
falloff of the multipole form factors at very large $Q^2$.

For very large $Q^2$ (pQCD regime), 
the transition amplitudes $G_\ell$ follow 
the power law falloffs from (\ref{eqGp-as})--(\ref{eqGm-as}).
The first consequence of the general analysis is that 
\ba 
& &
G_{E0}, \;  G_{M1} \propto   \frac{1}{Q^4},
\label{eqGE0-as}  \\
& &
G_{E2}, \; G_{M3} \propto   \frac{1}{Q^6}. 
\label{eqGE2-as} 
\ea

The analysis of the amplitude $G_-$, determined by $G_2$,
demonstrates that, in addition to (\ref{eqGE0-as}) 
and (\ref{eqGE2-as}), for  \mbox{$\tau \gg 1$}, 
one has also
\ba
G_{M1} \simeq \frac{4}{5} \tau G_{M3}. 
\label{eqGM1-infty}
\ea
The implication of the previous relation 
is that the two magnetic form factors 
are correlated for very  large $Q^2$.

The relations $G_{M1} \propto 1/Q^4$, $G_{M3} \propto 1/Q^6$ and
(\ref{eqGM1-infty}) imply that the leading-order 
the terms in $G_{M1}$ and $\tau G_{M3}$ are  
given by terms of order $1/Q^4$, which cancel exactly  
in $G_{M1} - \frac{4}{5} \tau G_{M3}$, 
and only the terms of higher order $1/Q^6$ survive.

To the best of our knowledge, 
the relation (\ref{eqGM1-infty}), valid for very large $Q^2$,
has not been discussed in the literature.
We emphasize, however, that the relations 
(\ref{eqGE0-as}),  (\ref{eqGE2-as}) and (\ref{eqGM1-infty})
are the consequence of the natural order 
of the amplitudes (\ref{eqGp})--(\ref{eqGm}).

The leading-order falloff of the form 
factors $F_i^\ast$ and $G_i$ ($i=1,..,4$) 
can also be estimated, based on the relations 
between representations (\ref{eqG1})--(\ref{eqG4}),  
(\ref{eqGE0-0})--(\ref{eqGM3-0}),  
(\ref{eqF1})--(\ref{eqF4}), and  
(\ref{eqG1a})--(\ref{eqG4a}).
Those results are presented in the first row (case 1) 
of Table~\ref{tableFF-largeQ2}.

\subsection{Alternative forms for the form factors falloffs}

The results (\ref{eqGE0-as}),  (\ref{eqGE2-as}) and 
(\ref{eqGM1-infty}) are then a consequence of the pQCD regime.
One can, nevertheless, look for weaker constraints at large $Q^2$.

If we remove the condition (\ref{eqGM1-infty}) 
but keep the conditions (\ref{eqGE0-as}) and (\ref{eqGE2-as}),
we obtain different falloffs for the form 
factors $F_i^\ast$ and $G_i$ ($i=1,..,4$).
The corresponding forms are presented in 
the second row (case 2) of Table~\ref{tableFF-largeQ2}.

The main consequence in the model which 
ignores the condition (\ref{eqGM1-infty}) 
appears in the amplitude $G_-$.
As before,  we obtain $G_+ \propto 1/Q^3$,
$G_{01}$, $G_{03} \propto 1/Q^4$.
In the case of $G_-$, since 
$G_- \propto Q G_2$ and $G_2 \propto 1/Q^4$
(instead of  $G_2 \propto 1/Q^6$), one obtains
\ba
G_- \propto \frac{1}{Q^3}, 
\ea
in contradiction with 
the asymptotic relation (\ref{eqGm-as}) 
expected from pQCD.

In summary, if we ignore the relation (\ref{eq-G2-extra}),
we fail to reproduce the natural order 
of the amplitudes:
$G_+ \propto 1/Q^3$, $G_{01}$, $G_{03} \propto 1/Q^4$,
and $G_- \propto 1/Q^5$.

\setcounter{table}{0}
\renewcommand{\thetable}{B\arabic{table}}

\section{Asymptotic relations for the covariant spectator quark model}
\label{appModel}

We discuss in this Appendix the asymptotic relations 
associated with the electromagnetic form factors of the $\Omega^-$
from Eqs.~(\ref{eqGE0})--(\ref{eqGM3}),
derived from the covariant spectator quark model framework.

\subsection{Uncorrelated form factors}

In a first step, we ignore the constraint (\ref{eqGM3extra}).
From the discussion from Sec.~\ref{secLargeQ2},
we know already that
\ba
{\cal I}_S \propto\frac{1}{Q^4}, \hspace{.5cm}
{\cal I}_{D3} \propto \frac{1}{Q^4}, \hspace{.5cm}
{\cal I}_{D1} \propto \frac{1}{Q^4}, 
\label{eqIntApp}
\ea
apart logarithmic corrections.

For the discussion of the large-$Q^2$ region, 
it is important to notice that 
the functions $\tilde g_\Omega$ and $\tilde f_\Omega$,
related with the strange quark form factors
$f_{10}(Q^2)$ and $f_{20}(Q^2)$,
behave as constants for very large $Q^2$,
\ba
& &
\tilde g_\Omega  \to  \kappa_q^\prime - \lambda_q, 
\label{eq-gtilP}
\\
& & 
 \tilde f_\Omega  \to - \lambda_q, 
\label{eq-ftilP}
\ea
where 
$\kappa_q^\prime =
\kappa_s \left[
d_0 \frac{m_\phi^2}{4 M_N M_\Omega} + (1-d_0) \frac{M_h^2}{4 M_N M_\Omega}
\right] $.
All parameters are defined in the 
strange quark form factors.
From our calibration to the strange quark current~\cite{Omega},
one has $\lambda_q = 1.21$ and  $\kappa_q^\prime =1.84$.

From Eqs.~(\ref{eq-gtilP}) and (\ref{eq-ftilP}), 
one can conclude that for large $Q^2$,
$\tilde f_\Omega$ is negative, and $\tilde g_\Omega$ is positive.
At low $Q^2$, these functions are both negative.

The combination of the results (\ref{eqIntApp}) 
with Eqs.~(\ref{eqGE0})--(\ref{eqGM3})  leads to 
the $G_{E0}$, $G_{M1} \propto 1/Q^4$ and  $G_{E2}$, $G_{M3} \propto 1/Q^6$.

We now look for the result for 
$G_{M1} - \frac{4}{5}\tau G_{M3}$.
From Eqs.~(\ref{eqGM1}) and (\ref{eqGM3}), 
one obtains
\ba
& &
G_{M1} - \frac{4}{5}\tau G_{M3} =
N^2 \tilde f_\Omega
 \left[ {\cal I}_S - 2 b\, {\cal I}_{D1} \right]. 
\label{eq-cond2}
\ea
From this result, we conclude 
that the relation between the two 
magnetic form factors is independent from the $D3$ component.

If the functions ${\cal I}_S$ and $b {\cal I}_{D1}$
are uncorrelated, we can conclude that 
$G_{M1} - \frac{4}{5}\tau G_{M3} \propto 1/Q^4$.
Working backward, we can derive the corresponding 
falloff for the form factors $G_i$ and $F_i^\ast$ ($i=1,2,3,4$)
and obtain the relations associated with the case 2 
from Table~\ref{tableFF-largeQ2},
discussed in Appendix~\ref{app-PQCD}.

\subsection{Asymptotic relations consistent with pQCD}

We consider now the possibility of imposing 
that the results of the covariant spectator quark model 
are consistent with the relation (\ref{eqGM3extra})
for very large $Q^2$.
From  (\ref{eq-cond2}), we conclude that 
the condition  (\ref{eqGM3extra}) is valid if 
\ba
& &
{\cal I}_S = 2 b\, {\cal I}_{D1},
\label{eq-cond2b}
\ea
apart terms of the order of $1/Q^6$,
since it was established already that the overlap 
integrals are of the order of $1/Q^4$.

The compatibility of the covariant spectator quark model with pQCD 
requires then that ${\cal I}_S = b\, {\cal I}_{D1}$.
The connection with between $b\, {\cal I}_{D1}$ and 
${\cal I}_S$ can be imposed numerically.

Combining the original expressions 
(\ref{eqGE0})--(\ref{eqGM3}) for the 
form factors with the condition (\ref{eq-cond2b}),
we can write for large $Q^2$
\ba
G_{E0} &=& N^2 \tilde g_\Omega \, {\cal I}_S, \\
\tau G_{E2} &=& N^2 \tilde g_\Omega \, a \,  {\cal I}_{D3}, \\
G_{M1} &=& \frac{4}{5} N^2 \tilde f_\Omega 
\left[{\cal I}_S +  a \, {\cal I}_{D3}  \right], \\
\tau G_{M3} &=& 
N^2 \tilde f_\Omega 
\left[{\cal I}_S +  a \,  {\cal I}_{D3}  \right].
\ea



\end{document}